\documentclass[a4paper,11pt]{article}
\pdfoutput=1 

\usepackage{jcappub} 

\usepackage[T1]{fontenc} 

\usepackage[normalem]{ulem}
\usepackage{array,multirow,graphicx}
\usepackage{dcolumn}
\usepackage{txfonts}
\usepackage{newlfont}
\usepackage{times}
\usepackage{bm}
\usepackage{hyperref}
\usepackage[figtopcap]{subfigure}
\usepackage{float}
\usepackage{ulem}
\usepackage{xcolor}
\usepackage[T1]{fontenc}
\usepackage{ae,aecompl}
\usepackage{nccmath}
\usepackage{soul}

\usepackage{xcolor}
\usepackage{xspace}
\usepackage{xfrac}
\usepackage{fontawesome}

\newcommand{\LCDM}{\Lambda \textrm{CDM}}
\newcommand{\fTCDM}{f(T)\text{-}\textrm{CDM}}

\graphicspath{{./figs/}}

\title{Toward a concordance teleparallel Cosmology II: Linear perturbation}

\author{Mahmoud Hashim,}%
\author{Amr A. El-Zant,}%
\author[1]{Waleed El Hanafy,\note{Corresponding author.}}%
\author{and Alexey Golovnev}%

\affiliation{Centre for Theoretical Physics, The British University in Egypt, P.O. Box 43, El Sherouk City, Cairo 11837, Egypt}

\emailAdd{mahmoud@aims.ac.za}
\emailAdd{amr.elzant@bue.edu.eg}
\emailAdd{waleed.elhanafy@bue.edu.eg}
\emailAdd{alexey.golovnev@bue.edu.eg}

\abstract{%
Late time cosmic acceleration may be achieved by modifying gravity on large scales.
This should also have consequences on the evolution of perturbations.
We thus extend our study of exponential infrared $f(T)$ teleparallel gravity to examine the viability of the theory at the linear perturbation level,  evaluating the full CMB and matter power spectra. As the theory does not introduce extra free parameters, it  fits within the minimal six parameter space of standard $\LCDM$. Using Planck 2018 CMB (TT+TE+EE+lensing) alone,
best fits predict those parameters to be almost identical to $\LCDM$, with slightly smaller $\chi^2_{min}$. The resulting $H_0=72.24\pm 0.64$ km/s/Mpc, which ``practically"  alleviates the tension with local measurements, due to late time phantom behaviour. Inclusion of BAO data however reduces $H_0$, reflecting furthermore systematic deviations from data that are also present in supernova distances and the growth rate of structure (increasing the apparent tension in the latter case).  As the theory, unlike other viable $f(T)$ models, does not reduce to $\LCDM$ through extra free parameters, those conclusions are generic; applying to any modified gravity or dynamical dark energy with phantom behaviour. With best fit parameters, the present scenario produces a CMB spectrum almost identical to $\LCDM$, with slight deviation at low-multipole $\ell < 30$, where cosmic variance is large. The matter power spectrum is also quite close to $\LCDM$; with percent level scale free modifications affecting modes significantly smaller than the horizon, arising primarily from modified background evolution.  More significant deviations appear on larger scales, and may in principle distinguish modified gravity scenarios of the type studied here from dynamical dark energy.
}

\begin{document}
\maketitle
\section{Introduction}\label{Sec:Intro}
Almost a hundred years of observations have led to the $\LCDM$ universe, with a cosmological constant $\Lambda$ and a pressureless cold dark matter (CDM) dominating its contents. Only six free parameters --- baryon density $\Omega_{b}h^2$, cold dark matter density $\Omega_{c}h^2$, angular scale of acoustic peaks $\theta_{MC}$, amplitude of fluctuations $A_s$, spectral index of fluctuations $n_s$ and reionisation optical depth $\tau_{re}$ --- may be employed to provide precise fits to a wide range of different types of observations. The dark sector invoked remains nevertheless unidentified. Particularly obscure is the source of the cosmological constant, which introduces a dark energy components with constant pressure $p_{\Lambda}$ and density $\rho_{\Lambda}$ related by a fixed equation of state $w_{\Lambda}=p_{\Lambda}/\rho_{\Lambda}=-1$.

The elusive character of its dark sector notwithstanding, the
model is highly successful in describing cosmic expansion history and large scale
structure. Pending  problems remain on (sub) galactic scales at low redshifts~\cite{2017Galax,2017ARA&A},
and apparent difficulties more recently arose in relation to the preponderance
of early massive galaxies and the formation of their central black holes
(e.g. \cite{2017Natur, 2019Natur, 2020ApJ, 2020ApJF, 2021arXiv210301459N,2015Natur, 2019ConPh}).
Yet such issues may in principle be resolved though better understanding of
the complex baryonic physics of galaxy formation; or,
in the case of low redshift small scale problems,
a better understanding of the nature of dark matter;
or, more exotically, of the form of the small scale
power spectrum of primordial
perturbations~\cite{Kamlidd2000,SEZ2020}.

Another apparent crisis, not readily amenable to such remedies,
lies in the increasing tension between the values of the current Hubble parameter, $H_0$.
The value inferred from the cosmic microwave background (CMB) by Planck base-$\LCDM$ is $H_0=67.4 \pm 0.5$ km/s/Mpc \citep{Aghanim:2018eyx}.  On the other hand,  that locally measured --- e.g. by Riess et. al.~\citep{Riess:2020fzl}, hereafter R20, using the distance ladder  combined with best complementary sources of Cepheid calibration --- is $H_0=73.2 \pm 1.3$ km/s/Mpc, with 1.8\% precision. This is in 4.2$\sigma$ tension with Planck.  Furthermore, both tracks are supported by reinforcing  independent measurements, which finally lead to tensions at 4--6$\sigma$ for different combinations \citep{Verde:2019ivm,DiValentino:2021izs} (see also \citep{Riess:2020sih}). This is also another apparent discrepancy between the amplitude of matter fluctuation known as the $\sigma_8$ tension, or $S_8\equiv \sigma_8 \sqrt{\Omega_m/0.3}$ tension. The tension between $S_8$ values inferred by Planck base-$\LCDM$ and those measured through late universe data --- e.g., the Kilo Degree Survey \citep[hereafter KiDs-450]{Hildebrandt:2016iqg}, or KiDs-450 combined with the VISTA Kilo-Degree Infrared Galaxy Survey (VIKING) \citep[hereafter KV-450]{Hildebrandt:2018yau} --- is above 2$\sigma$.

In this context, it would seem that despite the successes of the $\LCDM$ paradigm, it may prove worthwhile to explore possible alternatives, including those invoking modified gravity as an alternative to the particularly enigmatic dark energy component.
In paper I \citep{Hashim:2020sez}, we examined the viability of the exponential IR teleparallel gravity $f(T)$,
where the torsion scalar $T$ acts as an effective dark energy component, substituting for $\Lambda$.
The form $f(T)=T e^{\beta T_0/T}$, where $\beta$ is a dimensionless parameter, is particularly appealing
because the resulting theory does not introduce extra free parameters than those of the $\LCDM$ since the model parameter $\beta$ is completely determined by the current density parameters. This makes the theory at hand statistically comparable, on equal footing, with $\LCDM$. In the aforementioned paper, we confronted the theory with observations characterizing the history of background expansion
and linear matter linear perturbation at the quasi-static approximation limit. The results confirmed that the theory can describe the cosmic evolution with statistical success similar to $\LCDM$, although it explains the late time acceleration in a completely different way. The evolution it described was also compatible with larger values of CMB-inferred Hubble parameter.

In this study (hereafter, paper II), we extend the analysis of the exponential IR $f(T)$ gravity to the linear perturbation level, to infer full CMB and late-time matter power spectra, and derive associated constraints on the theory. The organization of the paper is as follows: In Sec. \ref{Sec:IRMG}, we give a brief presentation of the $f(T)$ cosmology focusing on the exponential IR $f(T)$ theory. In Sec. \ref{Sec:Linear-pert}, we present the basic equations of the cosmological linear perturbation in $f(T)$ gravity. In Sec. \ref{Sec:Data-results}, we review the observational datasets and the methodology adopted through this analysis.  Also, we present the best-fit of the six parameter space of $\LCDM$ and $\fTCDM$ using CMB data alone and in combination with BAO in addition to some derived parameters. We focus the discussion on the early and the late universe tensions along with a comparison with external datasets within the exponential IR $f(T)$ gravity. In Sec. \ref{sec:spectra}, we give the power spectra of the fluctuations and their evolution. In Sec. \ref{Sec:Conclusion} we outline our conclusions.
\section{The exponential infrared $\lowercase{f}(T)$ gravity}\label{Sec:IRMG}
We take the action of a generalized teleparallel gravity
\begin{equation}
    \mathcal{S}=\frac{1}{2\kappa^2}\int d^{4}x\, |e|f(T)+\mathcal{S}_{M},
    \label{action}
\end{equation}
where $T$ denotes the teleparallel torsion scalar, which differs from the Ricci scalar by a total derivative term; $|e|=\sqrt{-g}=\det\left({e}_\mu{^a}\right)$; $\kappa\equiv \sqrt{8\pi G}$, where $G$ is Newton's constant; and $\mathcal{S}_{M}$ is the action of standard matter $m$ and radiation $r$.
For more details regarding teleparallel geometry and $f(T)$ gravity we refer to  \citep{Aldrovandi:2013wha,Cai:2015emx,Krssak:2018ywd},

We assume that the spatial background geometry of the universe is of a flat Friedmann-Lema\^{\i}tre-Robertson-Walker (FLRW) form. Hence, we take the Cartesian coordinate system ($t;x,y,z$) and the diagonal vierbein
\begin{equation}
   {e_{\mu}}^{a}=\textmd{diag}\left(1,a(t),a(t),a(t)\right),
\label{tetrad}
\end{equation}
where $a(t)$ is the scale factor of the universe. The above vierbein generates the flat FLRW spacetime metric
\begin{equation}
   dS^2=-dt^{2}+a(t)^{2}\delta_{ij} dx^{i} dx^{j},
\label{FRW-metric}
\end{equation}
with Minkowskian signature is $\eta_{ab}=(-;+,+,+)$. This defines the teleparallel torsion scalar
\begin{equation}
T=6 H^2. \label{eq:Torsion_sc}
\end{equation}
The variation of the action \eqref{action} with respect to the vierbein gives rise to the field equations
\begin{equation}
    \frac{1}{\kappa^2_{\textrm eff}} \mathfrak{G}_{\mu\nu}= \mathfrak{T}^{(M)}_{\mu\nu}+\mathfrak{T}^{(DE)}_{\mu\nu},
\label{field_eqns}
\end{equation}
where $\kappa^2_{\textrm eff} = \kappa^2 /f_T$, $\mathfrak{T}^{(M)}_{\mu\nu}$ is the energy-momentum tensor of the matter sector. The $f(T)$ contribution to the energy-momentum can then be expressed as
\begin{equation}
\label{torsion-Tmn}
\mathfrak{T}^{(DE)}_{\mu\nu}=\frac{1}{\kappa^2} \left(\frac{1}{2}g_{\mu\nu}\left(Tf_T-f\right)-f_{TT}S_{\nu\mu\rho}\nabla^{\rho}T\right).
\end{equation}
In the context of modified gravity theories of the type tackled here, the dark energy like sector $\mathfrak{T}^{(DE)}_{\mu\nu}$ in the field equations \eqref{field_eqns} is sourced by geometrical terms
that mimic the effect of ``physical'' (scalar field sourced)  dark energy.

Motivated by the phase portrait patterns of viable models involving late acceleration phases
sourced by geometrical terms, an exponential IR $f(T)$ theory had been proposed \citep{Awad:2017yod}
\begin{equation}
f(T)=T e^{\beta \left(T_0/T\right)}, \label{exp-IR}
\end{equation}
where $T_0 = 6 H_0^2$ and $\beta$ is a dimensionless parameter. In its context, the Friedmann equation reads \citep{Hashim:2020sez}
\begin{equation}
\left(E^2-2\beta\right) e^{\frac{\beta}{E^2}}=\Omega_{m,0}(1+z)^3+\Omega_{r,0}(1+z)^4, \label{FR-E-exp-IR}
\end{equation}
where $E=H/H_0$, and $\Omega_{m,0}$ and $\Omega_{r,0}$ are the current values of the matter and the radiation density parameters, respectively.

Remarkably, at $z=0$ (and $E=1$), the $\beta$-parameter can be expressed in terms of the current density parameters,
\begin{equation}
    \beta=\frac{1}{2}+\mathcal{W}\left(\frac{\Omega_{m,0}+\Omega_{r,0}}{-2e^{\frac{1}{2}}}\right), \label{beta}
\end{equation}
where $\mathcal{W}(x)$ is the Lambert $\mathcal{W}$ function\footnote{defined via $x=\mathcal{W}(x) \, \exp{\mathcal{W}(x)}$.}.
With this relation comes a significant advantage;
as it implies that the model does not introduce any new parameters in addition to those in $\LCDM$. This, unlike other viable $f(T)$ theories \citep{Bengochea:2008gz,Linder:2010py,Bamba:2010wb}.
The analysis of the theory in paper I shows that the model is viable at the background level \citep{Hashim:2020sez}.  In the present study we examine in some
detail its predictions at the linear perturbation level.

We note that the GR limit is recovered by setting $\beta=0$.
Also, at earlier times, when Hubble values are large, the background
evolution described by the function $f(T)$, reduces to that described
by GR. In contrast, in the small Hubble regime, we expect  deviations from
GR at the background cosmological evolution level. As we will see here
this is also accompanied by significant deviations in the evolution of linear
perturbation on large (close to horizon) scales.
\section{Linear perturbations in $\lowercase{f}(T)$ gravity}
\label{Sec:Linear-pert}
We consider perturbation theory in conformal time, $a(\tau)\cdot d\tau=dt$, so that the background vierbein takes the form ${e_{\mu}}^{a}=a(\tau)\cdot \textmd{diag}\left(1,1,1,1\right).$

When studying perturbations one has to bear in mind that, on top of the ten independent components of the metric, a vierbein has six more components related with the local Lorentz rotations of the frame: an arbitrary boost with a scalar and a vector part, and an arbitrary rotation with a (pseudo)scalar and a (pseudo)vector part, too. However, the diffeomorphism symmetry is the same as in GR. And, precisely as in GR, we can set to zero two scalar quantities and one vector. (See Ref. \citep{Golovnev:2018wbh} for details.)
In particular, we can choose the Newtonian gauge for scalar perturbations, while tensor perturbations are gauge invariant. Note also that, much like in GR, the vector perturbations are not important  \citep{Golovnev:2018wbh}, and we will not discuss them in what follows.

In the Newtonian gauge the scalar metric perturbation takes the form
\begin{equation}
dS^2=a^2(\tau)\left(-(1+2\psi)d\tau^2+(1-2\phi)\delta_{ij}dx^i dx^j\right),
\end{equation}
with  the two potentials $\phi$ and $\psi$ describing the scalar modes of the metric perturbations, and $x_i$'s being comoving coordinates.

The most general linear scalar perturbation of the vierbein which gives this linear metric perturbation reads
\begin{eqnarray}
e^0_0 & = & a(\tau)\cdot\left(1+\psi\right)\\
e^0_i & = & a(\tau)\cdot \partial_i \zeta\\
e^a_0 & = & a(\tau)\cdot \partial_a \zeta\\
e^a_j & = & a(\tau)\cdot \left((1-\phi)\delta^a_j+\epsilon_{ajk}\partial_k s\right),
\end{eqnarray}
where $\zeta$ represents the scalar part of a Lorentz boost, and $s$ is the scalar part of a spatial rotation (vectorial parts are given in both cases by divergence-less vectors instead of the gradients). The choice of Newtonian gauge is already made here, so that the same $\zeta$ is in two different components and no $\partial\partial\sigma$ term in the spatial part.
The difference here with the formulas from the Ref.  \citep{Golovnev:2018wbh} is that we have fixed the Newtonian gauge from the very beginning, neglected the vector and tensor perturbations, and changed the names of the Newtonian potentials, $\phi\leftrightarrow\psi$, in order to be in accordance with the Ref. \citep{Ma:1995ey}.

One can easily see that $s$ does not contribute to the linearised equations of motions, and therefore, apart from the usual two potentials, we have one new variable in the scalar sector, $\zeta$.
The new variables are governed by the anti-symmetric part of equations. In our case it is identically satisfied in the spatial part leaving the pseudoscalar $s$ free, while from the mixed components we get:
\begin{equation}
\label{asymm}
k^2\zeta=3\left(\phi^{\prime}+{\cal H}\psi-\frac{{\cal H}^{\prime}-{\cal H}^2}{\cal H}\phi\right),
\end{equation}
where ${\cal H} = a H$, the dash denotes a derivative with respect to conformal time $\tau$, and $k$ is the wavenumber, so that $k^2\zeta=-\bigtriangleup\zeta$. In the following, we parameterize the deviations from the standard scenario in terms of two parameters $Q=1/f_{T}$ and $\Xi=12 (\mathcal{H}'-\mathcal{H}^2) Q f_{TT}$, whereas the GR limit is obtained as $Q\to 1$ and $\Xi \to 0$.

In the symmetric part the perturbation equations are analogous to the usual ones. The temporal components give
\begin{equation}
\label{eq:pert1}
k^2\phi+3{\cal H}(\phi^{\prime}+{\cal H}\psi)+\frac{3\mathcal{H}^2 \Xi}{a^2}\phi=-4\pi Q G a^2\delta\rho
\end{equation}
where $\delta\rho$ is the density fluctuation, and we have used the equation (\ref{asymm}) for $k^2\zeta$.

In the mixed components, the same procedure of eliminating $\zeta$ yields
\begin{equation}
\label{smix}
\phi' = 4\pi Q G \left(\frac{a}{k}\right)^2 (\rho + p)\theta - {\cal H}\psi - \frac{\mathcal{H} \Xi}{a^2}\phi
\end{equation}
where $\theta$ is divergence of the velocity of the matter fluid. In Ref. \citep{Golovnev:2018wbh} the velocity potential $u$ was used instead, the relation between the two being $\theta=-k^2 u$.

In the spatial part, the off-diagonal components give
\begin{equation}\label{eq:pert2}
\psi = \phi - 12 \pi Q G \left(\frac{a}{k}\right)^2 (\rho+p)\sigma - \mathcal{H} \Xi \,\zeta
\end{equation}
showing the presence of the gravitational slip even in absence of anisotropic stress $\sigma$. Note that using equations (\ref{asymm}, \ref{smix}) above $\zeta$ can be eliminated from the right hand side as
\begin{equation}\label{eq:grav_slip}
\zeta = \frac{3}{k^2} \left[\left(\frac{\mathcal{H}}{a^2}\left(a^2-\Xi\right)-\frac{\mathcal{H}'}{\mathcal{H}}\right) \phi + 4\pi Q G \left(\frac{a}{k}\right)^2 (\rho + p)\theta  \right].
\end{equation}

 In general, therefore, one expects a large gravitational slip to affect super-horizon scales within $f(T)$ gravity, unless the universe is at a quasi de Sitter phase. Notably, the gravitational slip $\zeta$ is sourced by a rescaling factor associated with the potential $\phi$ and an additive term associated with the divergence of the velocity of the matter fluid $\theta$. The first is $\propto k^{-2}$ and the second is $\propto k^{-4}$, the latter provides a significant relativistic correction near the observed horizon scales $r_H=c/H_0$ (i.e. $k \sim 2.4\times 10^{-4}$ Mpc$^{-1}$). This is consistent with IRMG scenario. It proves convenient to parameterize deviations from standard cosmology by substituting from Eqs \eqref{smix} and \eqref{eq:grav_slip} into \eqref{eq:pert1} and \eqref{eq:pert2}, respectively. Then, we write
\begin{eqnarray}
k^2 \phi &=& -4 \pi Q G a^2 \Delta \rho,\label{eq:Poisson-like}\\
k^2\left(\psi-R \phi\right) &=& -12\pi Q G a^2  (\rho+p)\sigma - 12 \pi Q G \Xi \left(\frac{a}{k}\right)^2 \mathcal{H}(\rho+p)\theta ,\label{eq:gslip}
\end{eqnarray}
where
\begin{eqnarray}
    \Delta \rho &=& \delta \rho + \frac{3\mathcal{H}(\rho+p)}{k^2}\theta,\\
    R &=& 1+\frac{3 \Xi}{k^2 a^2}\left[\mathcal{H}^2 \Xi + (\mathcal{H}'-\mathcal{H}^2) a^2 \right].
\end{eqnarray}
At the GR limit, clearly $Q\to 1$ and $\Xi\to 0$ (subsequently $R\to 1$). We note that Eqs. \eqref{eq:Poisson-like} and \eqref{eq:gslip} reduce to the generic parametrized equations given in \citep{Bertschinger:2008zb,2010PhRvD..81h3534B} (and used in \citep{Dossett:2011tn,2014JCAP...03..046D,Dossett:2015nda}), except for an extra relativistic correction in terms of $\theta$.

For the particular $f(T)$ theory at hand, which is almost indistinguishable from GR at large Hubble values (earlier times),
the effect of the modified evolution would be strongest on modes that enter the horizon late, in the context of the standard inflationary
scenario for the origin of cosmological perturbations --- that is on the largest observable scales at relatively low redshift.
As we will see in Section~\ref{Sec:Lin-pert-evol} the associated effects may be large, but detection would be hindered, for the present model, by cosmic variance.

The diagonal part of spatial components gives a somewhat cumbersome equation  containing the fluctuation of pressure $\delta p$, though it is redundant with other equations by virtue of Bianchi identities~\citep{Golovnev:2018wbh}.
Tensor perturbations are easy to derive. Their dynamics is very similar to GR (for perfect fluid matter):
$$h_{ij}^{\prime\prime}+2{\cal H}\left(1 + \frac{\Xi}{2a^2}\right)h_{ij}^{\prime}+k^2 h_{ij}=0,$$
the difference being due to modified background expansion.
An important point is that the speed of GW remains the same as in GR.

To complete our review of linear perturbation, we take the usual perfect fluid approximation to describe the matter content
\begin{equation}
\mathfrak{T}^{(M)\mu}{_\nu}=p g{^\mu}{_\nu}+ \left(\rho+p\right)U^{\mu}U_{\nu},
\end{equation}
where $\rho$, $p$ and $U^{\mu}=dx^{\mu}/\sqrt{-dS^2}$ are the density, pressure and 4-velocity unit vector of the fluid, respectively. As in the usual treatment, the continuity of the perturbed energy-momentum in $k$-space implies \citep{Ma:1995ey}
\begin{eqnarray}
\delta' &=& -(1+w)\left(\theta-3\phi'\right) + 3{\cal H}\left(w - \frac{\delta p}{\delta \rho}\right)\delta\,,\\
\theta' &=& -{\cal H}\left(1-3w\right)\theta - \frac{w'}{1+w}\theta + \frac{\delta p/\delta \rho}{1+w}k^2 \delta  - k^2 \sigma + k^2 \psi\qquad.
\end{eqnarray}
These equations are valid for a single uncoupled fluid.
\section{Data, Methodology and results}\label{Sec:Data-results}

A successful 'concordance' model should predict cosmological parameters consistent with late-time data even by considering early-time data alone, and vice versa. Here, we consider the full Planck 2018 legacy (hereafter PL18) data alone and the combined PL18 and BAO
data for both $\LCDM$ and $\fTCDM$.

To calculate the evolution of cosmological perturbations in
exponential IR $f(T)$ gravity, we use the  linear perturbation equations  in Section~\ref{Sec:Linear-pert}. The full CMB spectra, on which
the results here are based, are given in Section~\ref{sec:spectra} below,
where we also present and discuss the
expected late time matter power spectra, which turn out
to be only slightly modified, and in a virtually scale independent
way, except for close to horizon scales.

\subsection{Datasets}\label{Sec:Data}
Following PL18 notations \citep{Aghanim:2018eyx}, we have
\begin{itemize}
    \item [$\bullet$] \textbf{CMB}. We use the full temperature power spectrum of Planck 2018 legacy data \citep{Aghanim:2018eyx} adopting the baseline \verb"Plik" high-multipole likelihood which includes the high-$\ell$ multipole ($30 \leq \ell \leq 2508$) TT likelihood and the high-$\ell$ ($30 \leq \ell \leq 1996$) EE and TE likelihoods as well. We also consider the low-$\ell$ temperature \verb"Commander" likelihood as well as the low-$\ell$ \verb"SimAll" polarization likelihood in the low-$\ell$ range ($2 \leq \ell \leq 29$) EE likelihood. We refer to this combination as ``Base" to represent TT,TE,EE+lowE. Noting that we do not include low-$\ell$ TE correlations similarly to PL18 analysis.
    \item [$\bullet$] \textbf{CMB lensing}. In addition, we use the CMB lensing-potential power spectrum likelihood of Planck 2018 legacy data \citep{Aghanim:2018eyx}.
    \item [$\bullet$] \textbf{BAO}. Similar to Planck collaboration analysis \citep{Aghanim:2018eyx}, we use 6dFGS \citep{Beutler:2011MNRAS}, SDSS--MGS \citep{Ross:2014qpa} and BOSS--DR12 \citep{Alam:2016hwk}.
\end{itemize}
We note that we do not use $H_0$ prior from local measurements at any stage through out this analysis. This is in order to test {\it a priori} the possibility that late time evolution can restore a natural cosmological concordance between early and late universe measurements.
As the modifications to the matter power spectra
are small ( $\lesssim 1.5$\%) and nearly scale free at BAO scales and redshifts
(Section {Sec:Lin-pert-evol}), the
standard procedures used to infer distance measures from the BAO apply in practice.

Auxiliary datasets, related to supernovae distance measurements and redshift space distortions
are identified as they are used.
Definitions related to the distance measurements used are included
in Appendix~\ref{App:Definitions}.

\subsection{Parameter Space}\label{Sec:Para-space}
The analysis has been performed with the publicly available \verb"CLASS" code \citep{Blas_2011} together with \verb"MONTE PYTHON" \citep{Audren:2013} after proper modifications to calculate the CMB power spectrum within the exponential IR $f(T)$ scenario as provided in Section \ref{Sec:Linear-pert}. We used \verb"GetDist" python package \citep{Lewis:2019xzd} to run Monte Carlo Markov Chain samples getting the 2D contour plots of the different model parameters by considering the following dataset combinations: Base, Base+lensing and Base+lensing+BAO for both $\LCDM$ and the exponential IR $f(T)$ gravity.

Since the $f(T)$ theory at hand does not introduce any free parameters, the parameter space is fully described by the usual six parameters of the $\LCDM$ model. In particular, we have the following parameter space
\begin{equation}\label{Prameter-space}
    \mathcal{P}\equiv\left\{\Omega_{b}h^2,\, \Omega_{c}h^2,\, 100\,\theta_{s},\,
    \tau_{re},\, n_s,\, \log \left(10^{10} A_s\right) \right\}.
\end{equation}

Our baseline assumption is similar to the $\LCDM$ model with purely adiabatic scalar primordial perturbations with a power-law spectrum. We assume three neutrinos species, approximated as two massless states, a single massive neutrino of mass $m_\nu$ = 0.06 eV, we also assume that they are stream like non-interacting, relativistic particles at the time of recombination. The priors used for the six parameter space $\mathcal{P}$ are taken as flat priors as in PL18 \citep{Aghanim:2018eyx}.

\subsection{Results}\label{Sec:Results}
\begin{table*}
\caption{Planck constraints on the six parameters of the $\LCDM$ and the $\fTCDM$ models at $68\%$ CL. The minimum value of $\chi^2$ for each model with different datasets is given in the last row. Here `Base' represents TT,TE,EE+lowE joint dataset.}
\label{Table:Results}
    \centering
\resizebox{\textwidth}{!}{%
\begin{tabular} { l c c c c c c}
\hline\hline
\multirow{3}{*}{Parameter}& \multicolumn{3}{c}{$\LCDM$} & \multicolumn{3}{c}{$\fTCDM$} \\
\cline{2-7}
& Base & Base+lensing  & Base+lensing+BAO & Base & Base+lensing  & Base+lensing+BAO \\
& $68\%$ limits & $68\%$ limits & $68\%$ limits & $68\%$ limits & $68\%$ limits & $68\%$ limits \\
\hline
{$100\Omega_{b}h^2              \dotfill$} &$ 2.236^{+0.014}_{-0.016} $ & $ 2.238\pm 0.015 $ & $ 2.244\pm 0.013 $ & $ 2.238\pm 0.015 $ & $ 2.241\pm 0.015 $ & $ 2.231\pm 0.013 $  \\[5pt]
{$\Omega_{c}h^2    \dotfill$} & $ 0.1202\pm 0.0014 $ & $ 0.1200\pm 0.0012 $ & $ 0.11934\pm 0.00092 $ & $ 0.1200\pm 0.0014 $ & $ 0.1196\pm 0.0012 $ & $ 0.12106\pm 0.00092 $ \\[5pt]
{$ 100\theta_{\textrm s} \dotfill$} & $ 1.04188\pm 0.00029 $ & $ 1.04190\pm 0.00030 $ & $ 1.04197\pm 0.00028 $ & $ 1.04189\pm 0.00030 $ & $ 1.04191\pm 0.00030 $ & $ 1.04179\pm 0.00029 $ \\[5pt]
{$\tau_{re}       \dotfill$} & $ 0.0542\pm 0.0077 $ & $ 0.0545\pm 0.0073 $ & $ 0.0562\pm 0.0074 $ & $ 0.0540\pm 0.0078 $ & $ 0.0535\pm 0.0074 $ & $ 0.0500\pm 0.0069 $  \\[5pt]
{$\ln{\left(A_s10^{10}\right)}       \dotfill$} & $ 3.045\pm 0.016 $ & $ 3.045\pm 0.014 $ & $ 3.048\pm 0.015 $ & $ 3.044\pm 0.016 $ & $ 3.043\pm 0.014 $ & $ 3.038\pm 0.013 $ \\[5pt]
{$n_s       \dotfill$} & $ 0.9650\pm 0.0044 $ & $ 0.9658\pm 0.0041 $ & $ 0.9673\pm 0.0037 $ & $ 0.9661\pm 0.0044 $ & $ 0.9668\pm 0.0042 $ & $ 0.9636\pm 0.0037 $ \\
\hline
{$H_0    $ [km/s/Mpc]}$\dotfill$& $ 67.31^{+0.57}_{-0.65} $ & $ 67.41\pm 0.54 $ & $ 67.72\pm 0.42 $ & $ 72.03\pm 0.70 $ & $ 72.24\pm 0.64 $ & $ 71.49\pm 0.47 $ \\[5pt]
{$\Omega_{m}   \dotfill$}& $ 0.3162\pm 0.0085 $ & $ 0.3149\pm 0.0074 $ & $ 0.3107\pm 0.0056 $ & $ 0.2758\pm 0.0078 $ & $ 0.2735\pm 0.0069 $ & $ 0.2818\pm 0.0053 $  \\[5pt]
{$\sigma_{8}    \dotfill$}& $ 0.8117\pm 0.0074 $ & $ 0.8116\pm 0.0059 $ & $ 0.8108\pm 0.0060 $ & $ 0.8425\pm 0.0075 $ & $ 0.8412\pm 0.0061 $ & $ 0.8433\pm 0.0058 $ \\[5pt]
{$S_8=\sigma_8(\Omega_m/0.3)^{0.5}              \dotfill$}& $ 0.833\pm 0.016 $ & $ 0.831\pm 0.013 $ & $ 0.825\pm 0.010 $ & $ 0.808\pm 0.016 $ & $ 0.803\pm 0.013 $ & $ 0.817\pm 0.010 $ \\[5pt]
{$z_{re} \dotfill$}& $ 7.66\pm 0.78 $ & $ 7.69\pm 0.73 $ & $ 7.85\pm 0.73 $ & $ 7.62\pm 0.79 $ & $ 7.56\pm 0.75 $ & $ 7.24^{+0.76}_{-0.68} $ \\[5pt]
{Age[Gyr]}$\dotfill$&  $ 13.796^{+0.026}_{-0.022} $ & $ 13.793^{+0.025}_{-0.022} $ & $ 13.782\pm 0.020 $ & $ 13.706\pm 0.026 $ & $ 13.699\pm 0.025 $ & $ 13.723\pm 0.020 $  \\[5pt]
{$z_{s} \dotfill$}& $ 1088.91^{+0.23}_{-0.21} $ & $ 1088.88\pm 0.21 $ & $ 1088.78\pm 0.17 $ & $ 1088.88\pm 0.22 $ & $ 1088.82\pm 0.21 $ & $ 1089.03\pm 0.17 $ \\[5pt]
{$r_s[{\textrm Mpc}] \dotfill$}& $ 144.47\pm 0.30 $ & $ 144.51\pm 0.26 $ & $ 144.64\pm 0.21 $ & $ 144.51\pm 0.30 $ & $ 144.59\pm 0.27 $ & $ 144.30\pm 0.22 $ \\[5pt]
{$z_{drag} \dotfill$}&  $ 1059.98\pm 0.30 $ & $ 1060.01\pm 0.31 $ & $ 1060.08\pm 0.29 $ & $ 1059.99\pm 0.30 $ & $ 1060.04\pm 0.30 $ & $ 1059.91\pm 0.29 $\\[5pt]
{$r_{drag}[{\textrm Mpc}] \dotfill$}& $ 147.04\pm 0.30 $ & $ 147.08\pm 0.26 $ & $ 147.19\pm 0.23 $ & $ 147.08\pm 0.30 $ & $ 147.14\pm 0.27 $ & $ 146.88\pm 0.23 $ \\[5pt]
\hline
{ $\chi^2_{min} $} & $ 1386.83 $ & $ 1389.91 $ & $ 1392.83 $ & $ 1386.69 $ & $ 1390.67 $ & $ 1397.25 $ \\
\hline\hline
\end{tabular}
}
\end{table*}

We have calculated the predicted values of the various parameters predictions at $68\%$ confidence level (CL)
for the $\LCDM$ and the $f(T)$ models using the dataset combinations described above Since the the models entail the same number of parameters, the statistical comparison is direct. We display the results in Table \ref{Table:Results}, using Base, Base+lensing and Base+Lensing+BAO.
As can be inferred, he $f(T)$ theory at hand predicts the six parameters with values quite similar to those associated with $\LCDM$.
The best values for $H_0$ however are systematically larger, especially when CMB data alone is invoked in the claculation.
This can be clearly seen in Fig. \ref{Fig:2DPlanck} which shows the 2D joint contours considering several combinations of the six parameters at 68\% and 95\% CL for $\LCDM$ and $f(T)$ gravity.

\subsubsection{The Hubble tension and associated systematics}\label{Sec:Hubble-para}

\begin{figure*}
    \centering
    \includegraphics[width=\textwidth]{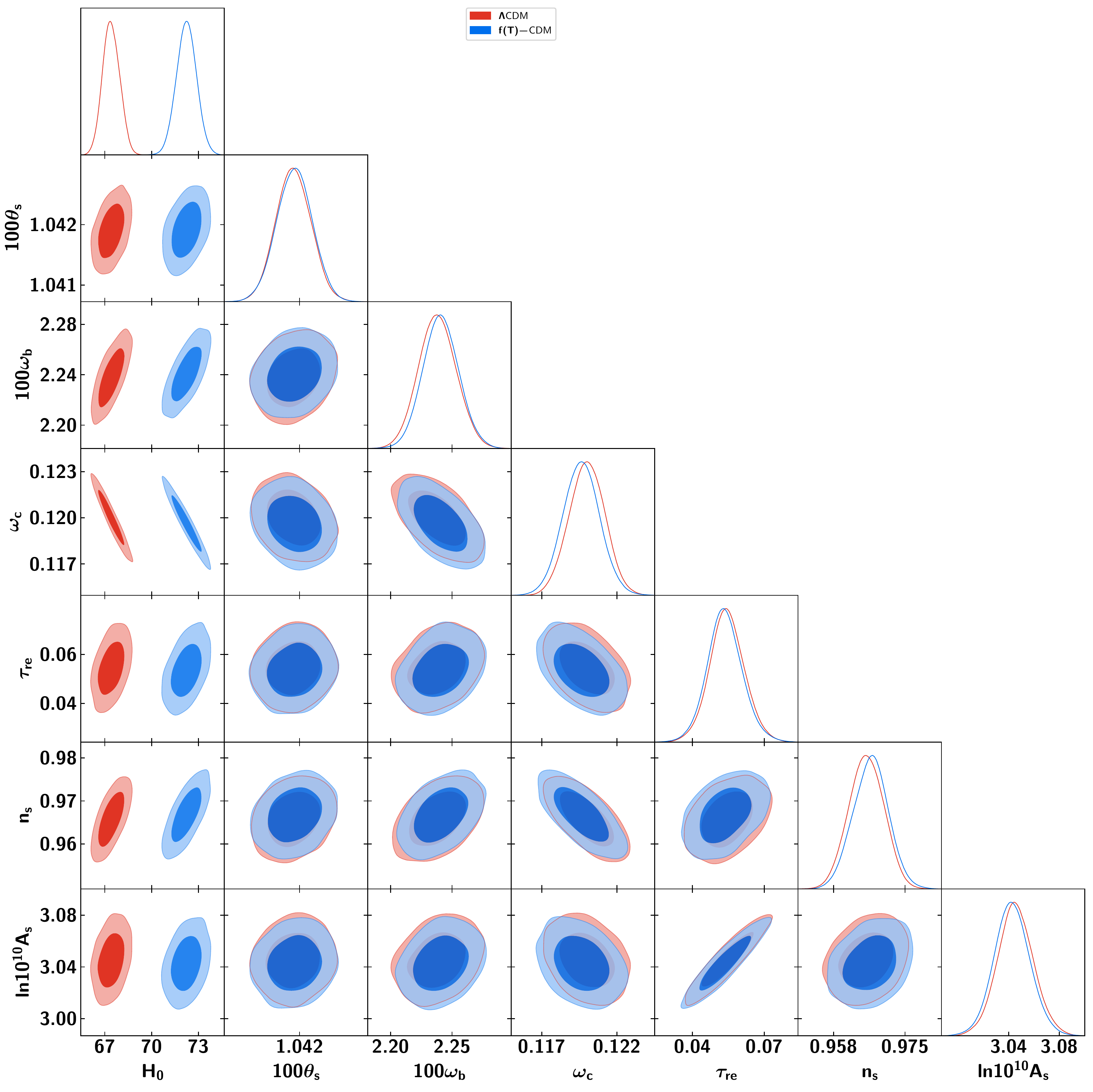}
    \caption{{The CMB constraints of $\LCDM$ and $\fTCDM$: 1D marginalized posterior distributions of the six parameter space in addition to the $H_0$ parameter of the $\LCDM$ and the exponential IR $f(T)$ theory together with the 2D joint contours at 68\% and 95\% CL for CMB alone}.}
    \label{Fig:2DPlanck}
\end{figure*}

Using CMB alone one can see from Fig.~\ref{Fig:2DPlanck} that
the $f(T)$ theory predicts a Hubble constant $H_0=72.03 \pm 0.70$ and $72.24 \pm 0.64$ km/s/Mpc with minimum chi-square $\chi^2=1386.69$ and $1390.67$, using Base and Base+lensing data combination, respectively.
In contrast to $\LCDM$ which predicts corresponding  values of $H_0=67.31^{+0.57}_{-0.65}$ and $67.41 \pm 0.54$ km/s/Mpc with chi-square $\chi^2=1386.83$  and $1389.91$. The upward  shift in mean values of $H_0$ in the $f(T)$ theory, rather broadening the uncertainties, is unlike the case with the models which extend base-$\LCDM$ by imposing extra free parameters to address the $H_0$ tension (cf. \citep{Vagnozzi:2019ezj}). The age of the universe, within the $f(T)$ framework, is $\sim 13.7$ Gyr (Table \ref{Table:Results}), which is consistent with astronomical data white dwarfs and
globular clusters. We further note that since the predictions of the suggested gravitational theory are expected to be virtually indistinguishable
from GR at early times, the sound horizon $r_{drag}$ is just as predicted by Planck base-$\LCDM$. The other successes of the standard model, such as a
baryon density consistent with big bang nucleosynthesis, are also retained.

\begin{figure}[t]
    \centering
    \includegraphics[width=0.48\textwidth]{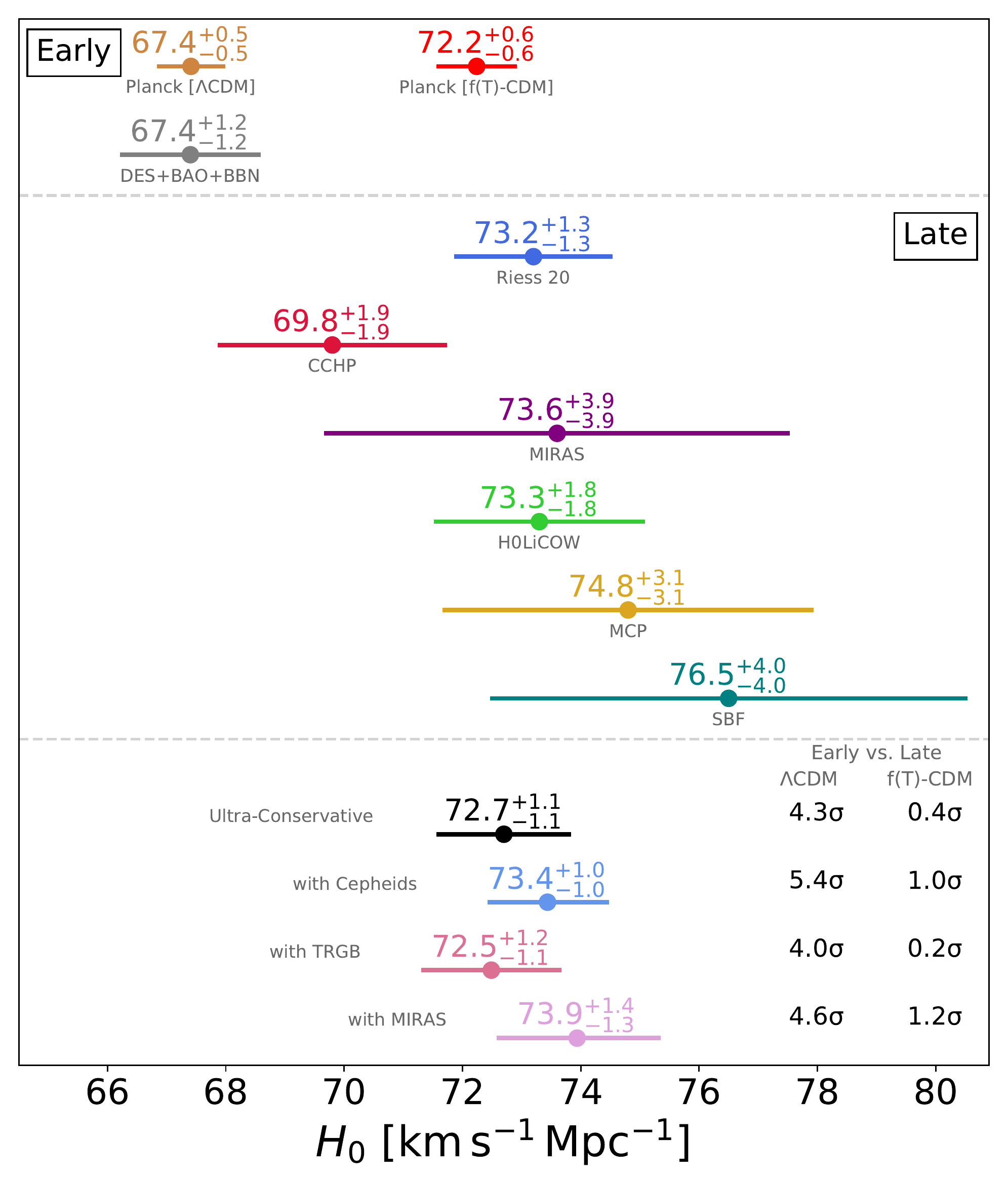}
     \includegraphics[width=0.48\textwidth]{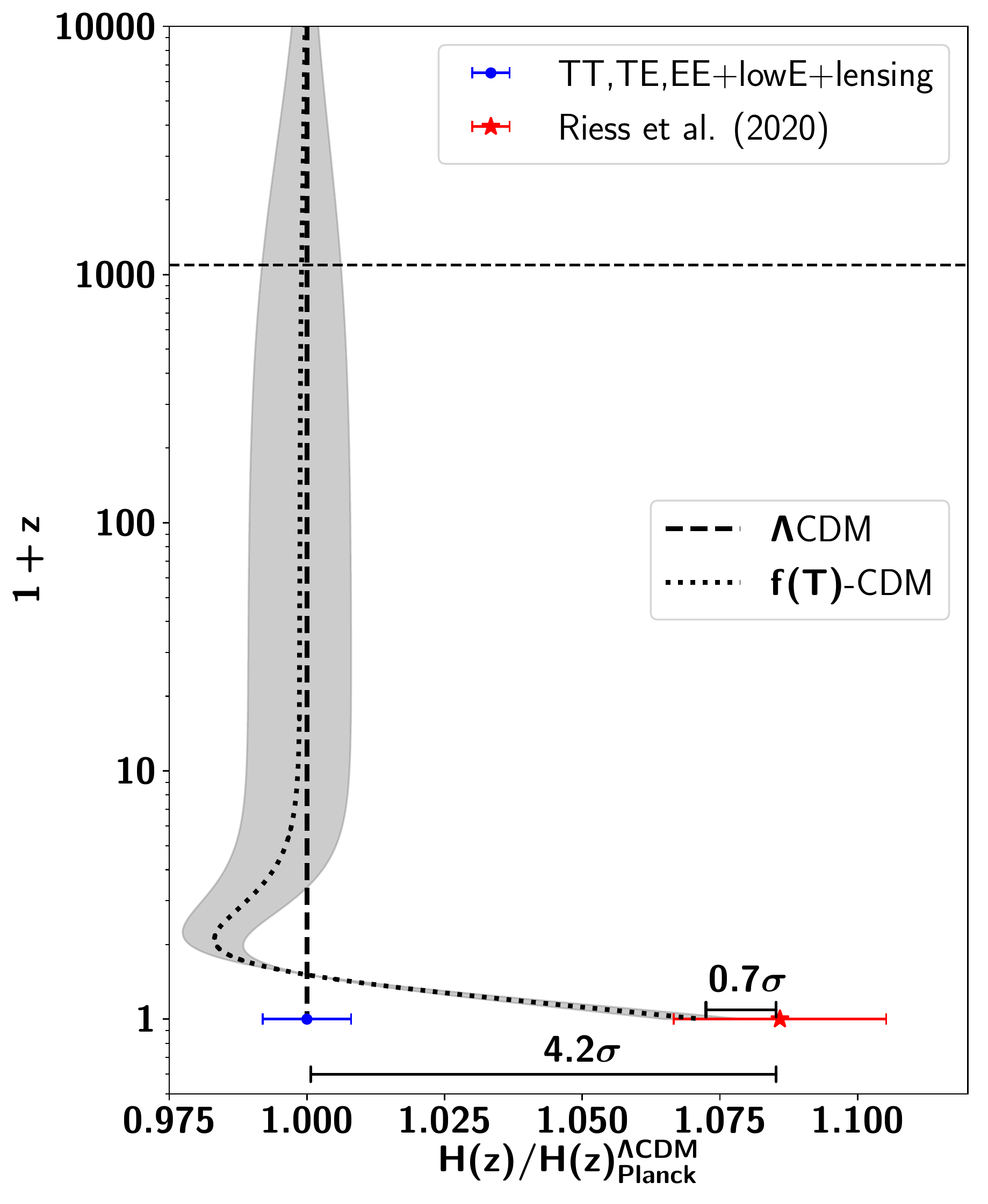}
    \caption{{Early versus late: 68\% CL constraint on $H_0$ from different cosmological probes. The left panel shows shows  early universe measurements of the Hubble constant, as inferred by Planck for base-$\LCDM$ (supported by an independent measurement using BAO+BBN+DES) and for base-$\fTCDM$ (using Planck CMB data); below the first dashed lines are late universe measurements of $H_0$; below the second dahsed lines are the tensions between late universe measurements of $H_0$ and Planck results of base-$\LCDM$ and base-$\fTCDM$, assuming linear error propagation by taking Gaussian approximations to the posterior distribution functions of each $H_0$ measurement. The ultra-conservative estimate $H_0=72.7 \pm 1.1$ km/s/Mpc as is obtained by \citep{DiValentino:2020vnx}. Right panel: Planck CMB constraints on the ratio of the Hubble rate $H(z)$ with respect to base-$\LCDM$ model, where the grey band represents 1$\sigma$ error of $H(z)$ according to Planck base-$\fTCDM$.}}
    \label{Fig:H0-tension}
\end{figure}

The above confirms that, seen simply as a contradiction between
CMB inferred and direct late time measurements of the Hubble constant,
the apparent $H_0$ tension can be effectively resolved by invoking a dark energy component involving a phantom regime.
The present formulation, in terms of modified gravity,
has the advantage of avoiding technical issues related to ghosts
that need to be dealt with when phantom regimes are achieved in terms of a scalar field component~\citep{El-Zant:2018bsc,Hashim:2020sez}.
In Fig.~\ref{Fig:H0-tension}, we summarize the tensions between early and late universe measurements with different combinations of late universe datasets, together with the Planck full likelihood results for $\LCDM$ as well as $\fTCDM$.
Notably, the combination of R20+H0LiCOW measures $H_0 = 73.2\pm 1.1$ km/s/Mpc, which is 4.8$\sigma$ away from Planck base-$\LCDM$ is indeed just 0.8$\sigma$ away from Planck base-$\fTCDM$.
The parameter space of the present theory involves six parameters, as in $\LCDM$,
and $\chi^2$ values are comparable for both models.

Nevertheless,  introducing BAO measurements changes the picture;
with the combination Base+lensing+BAO, one obtains a lower value of
$H_0=71.49\pm 0.47$ km/s/Mpc (with $\chi^2=1397.25$) with the best fitting $f(T)$ model.
This is more in line with results previously obtained in paper I \citep{Hashim:2020sez}, which gave $ H_0=70.52\pm 0.71$ km/s/Mpc for the combination $H_0$+SNIa+BBN+BAO+RSD+CMB $\theta_s$. It is consistent at 68\% CL with our findings here using Planck CMB (TT,TE,EE+lowE+lensing)+BAO. The latter combination gives more
statistical weight to CMB data, which may in turn be behind the somewhat larger $H_0$ value.

\begin{figure*}
    \centering
    \includegraphics[width=0.49\textwidth]{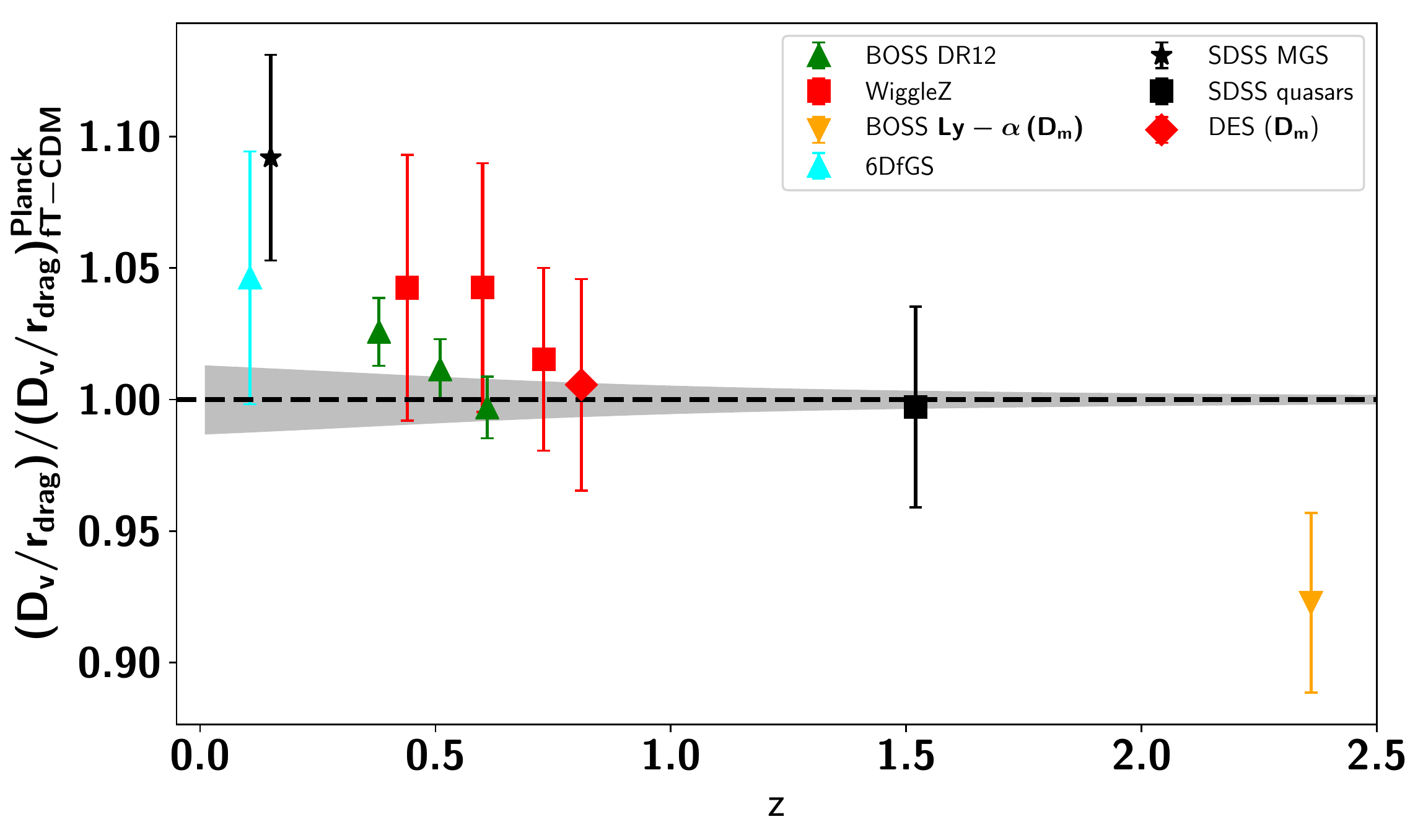}
    \includegraphics[width=0.49\textwidth]{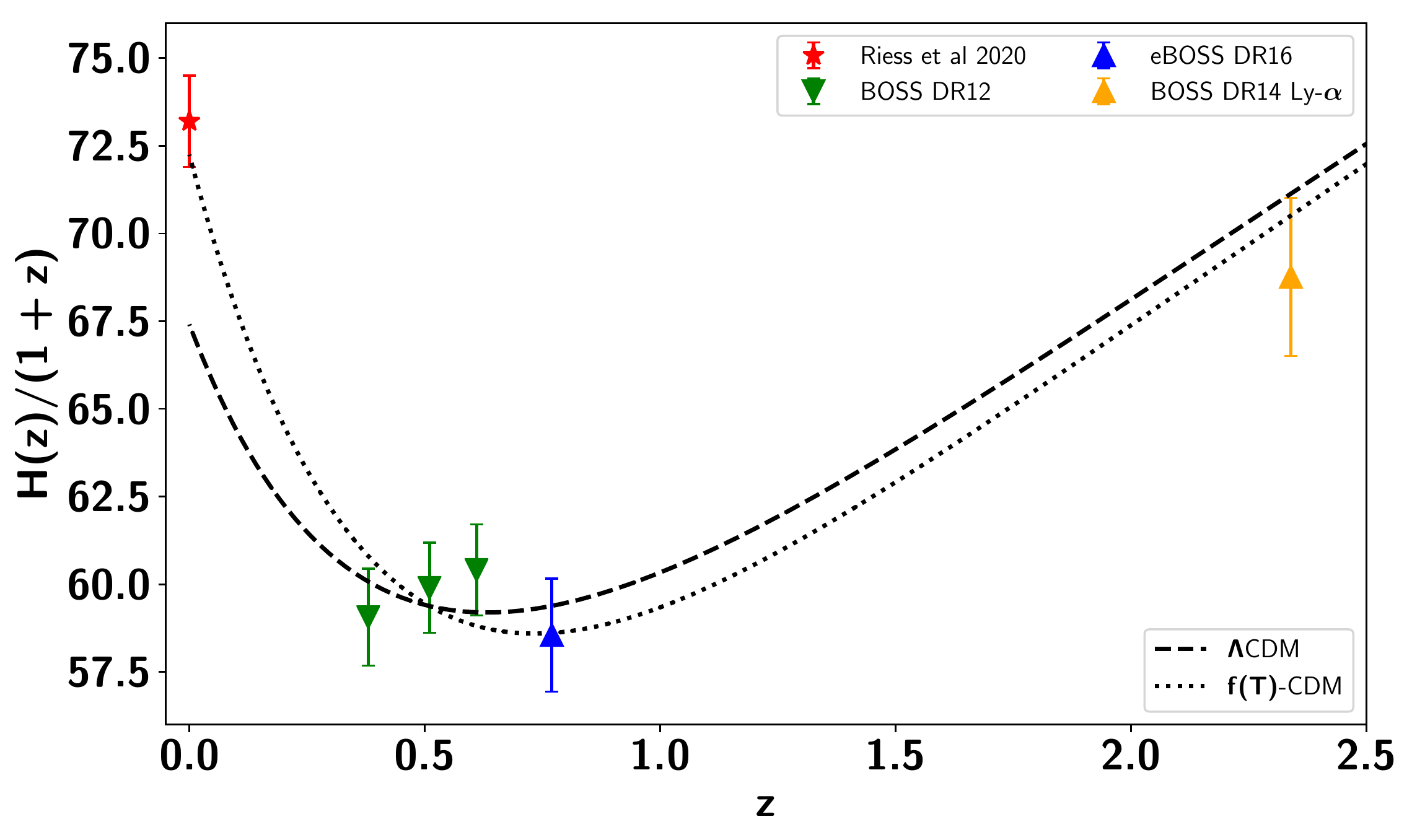}\\
    \includegraphics[width=0.53\textwidth]{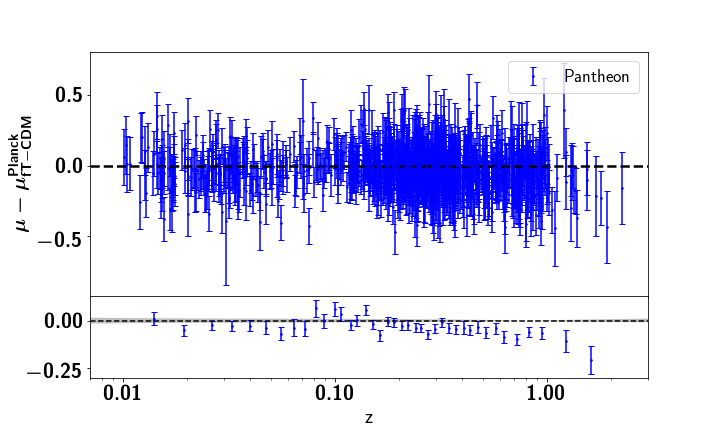}
    \includegraphics[width=0.45\textwidth]{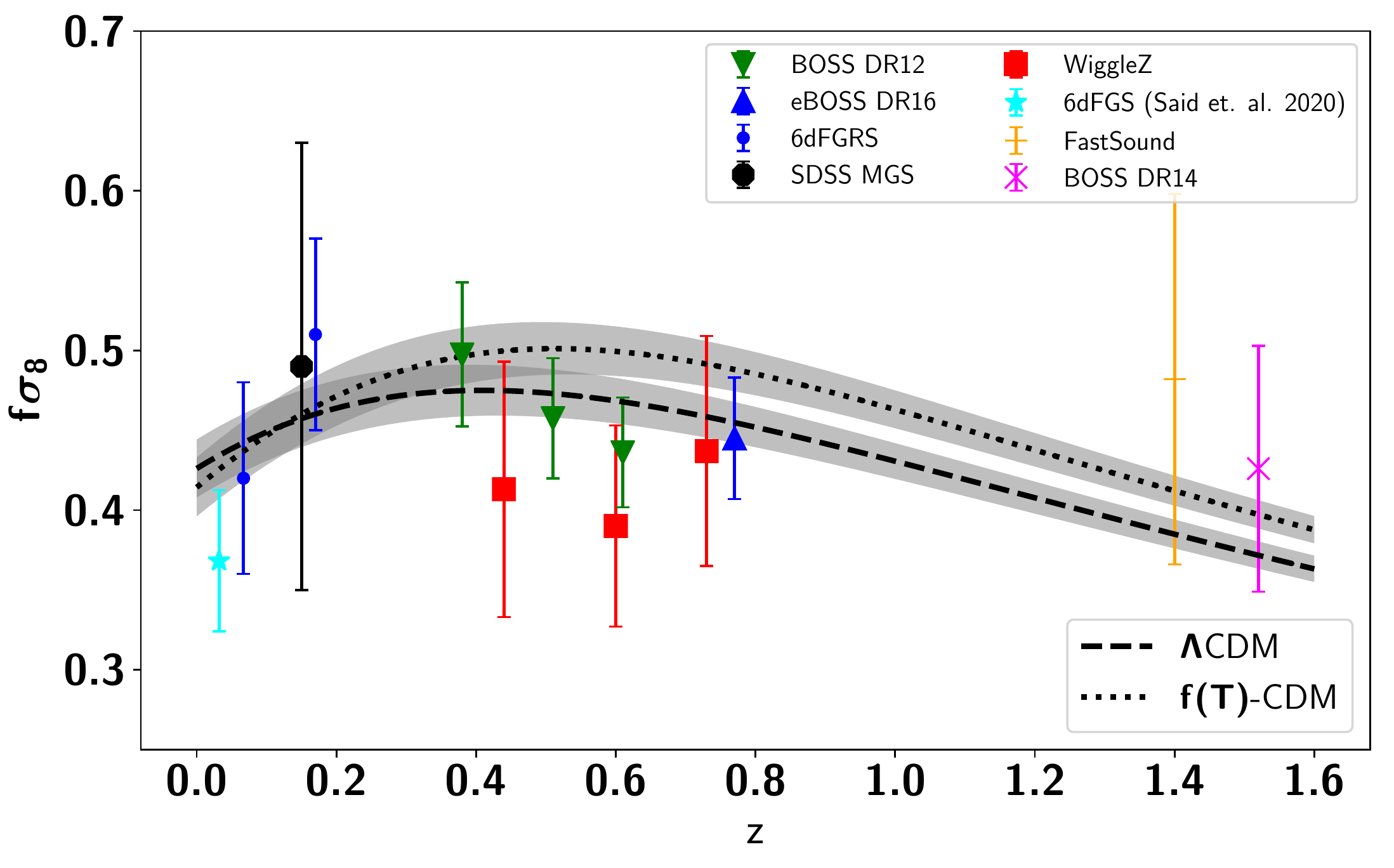}
    \caption{{Upper Left: Acoustic-scale measurements of the ratio $D_v/r_{drag}$
    divided by the corresponding mean distance ratio from Planck CMB (Base+lensing) base-$\fTCDM$ theory (definitions of distances used are given in Appendix~\ref{App:Definitions}).
    Upper Right: We compare the Hubble evolution of $\LCDM$ and $\fTCDM$ according to the best-fit values as given in Table \ref{Table:Results} together with radial BAO measurements BOSS DR12, eBOSS DR16 and BOSS DR14 Ly-$\alpha$ auto-correlation at $z$ = 2.34 \citep{Agathe:2019vsu}. Lower Left: A comparison of the distance modulus of Pantheon data \citep{Scolnic:2017caz} to the theoretical predictions of the Planck TT,TE,EE+lowE+lensing combined data of $f(T)$ theory best-fit. Lower Right: We compare the growth rate of fluctuations of $\LCDM$ and $\fTCDM$ according to the best-fit values of the parameters given in Table \ref{Table:Results} together with observational constraints from various redshift surveys from BOSS DR12 results \citep{Ross:2016gvb}, eBOSS DR16 \citep{Wang:2020tje}, 6dFGRS \citep{Beutler:2011MNRAS}, SDSS MGS \citep{Howlett:2014opa}, WiggleZ \citep{2012MNRAS.425..405B}, the growth rate constraint by \citep{Said:2020epb}, FastSound \citep{Okumura:2015lvp} and BOSS DR14 quassars \citep{Zarrouk:2018vwy}. Grey bands represent the 1$\sigma$ range allowed by Planck Base+lensing}.}
    \label{Fig:ext_data}
\end{figure*}

Furthermore, the deviations that result in smaller $H_0$ value when the BAO data are included reflect a systematic mismatch with late universe evolution associated with the phantom regime invoked by the present $f(T)$ model.
This effect, which can't be simply inferred by comparing $\chi^2$ values,
can be clearly observed in  Fig.~\ref{Fig:ext_data}; in particular, if the the cosmological model
is fixed through the CMB measurements, the volume averaged comoving angular diameter distance, $D_V$
systematically deviate from the measured values in the case of $f(T)$ (which is not the case with the standard model
\citep{Aghanim:2018eyx}).

The distance modulus of SNIa, $\mu(z)$, also shows a similar effect.
Indeed, including Pantheon SNIa data \citep{Scolnic:2017caz}, thus taking the combination Base+lensing+BAO+SNIa, we obtain $H_0(\LCDM)= 67.83\pm 0.57$ km/s/Mpc with $\chi^2= 1709.81$ and $H_0(\fTCDM)= 71.51\pm 0.60$ km/s/Mpc with $\chi^2=1718.75$. This is virtually indistinguishable from the corresponding
result without the SNIa but including the BAO.

The same systematic discrepancy in
the redshift space distortion (RSD)
measurements of the rate of the growth of structure\footnote{The definition of the growth rate is given in paper I \citep[Sec. 2.2]{Hashim:2020sez}} $f(z)\sigma_8$,
also shows similar deviations.
For the growth rate $f\sigma_8$  the systematics appear in both $f(T)$ and $\LCDM$ cosmologies but are more severe in the $f(T)$ case. In Appendix~\ref{App:S8-tension} we discuss cosmic shear data that helps break the degeneracy in $f \sigma_8$ at $z =0$. The results are
similar for both models for $\sigma_8$, with a slight amelioration in the $S_8 = \sigma_8 \sqrt{\Omega_m/0.3}$ tension. One notes however that as
$z \rightarrow 0$ our $f(T)$ based model dynamics is already approaching a cosmology with a cosmological
constant (Paper I, Fig.~1).

The phenomenon just discussed is generic: Attempting to resolve the tension between CMB-inferred  and late time measured $H_0$
through modifying late time cosmic expansion, particularly by invoking a phantom stage, necessarily entails tension with
BAO-based measurements~\citep{El-Zant:2018bsc}. The modified gravity model at hand introduces a phantom-like regime only relatively late (see the right panel of Fig. \ref{Fig:H0-tension}), which keeps the tension with BAO measurements at lesser level than in the aforementioned work, but the same phenomenon is at work.  Furthermore, as the exponential IR $f(T)$-based model used here has the same number of free parameters as $\LCDM$,  it does not reduce to $\LCDM$ as other viable $f(T)$ theories through particular choices of extra free parameters. Therefore, the conclusions presented here inevitably exist for any modified gravity or dynamical dark energy with phantom behaviour.


\section{Power spectra of linear perturbations}
\label{sec:spectra}

\subsection{CMB temperature and polarisation power spectra}\label{Sec:TT+TE+EE}
\begin{figure*}
    \centering
    \includegraphics[width=\textwidth]{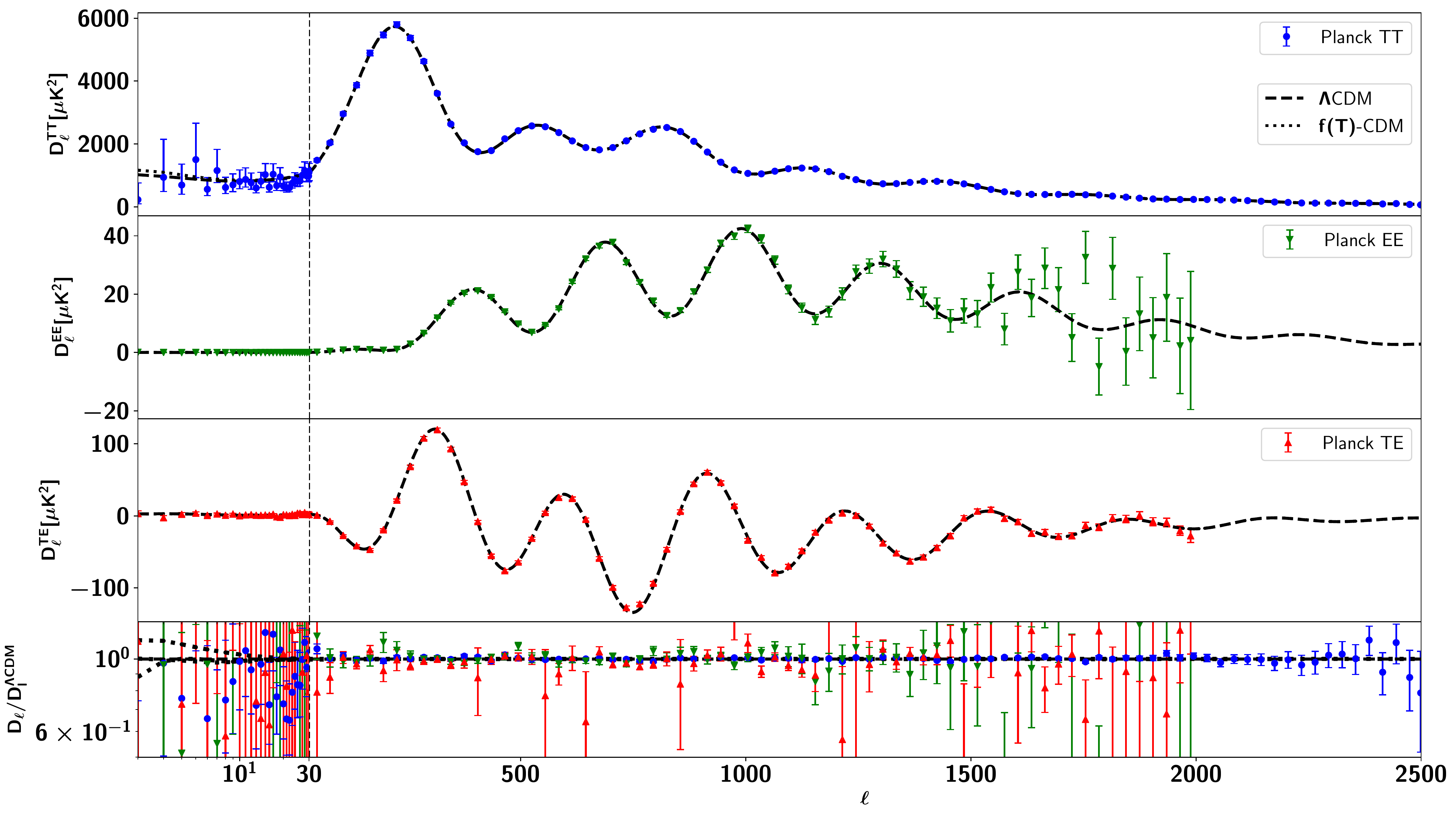}
    \caption{{CMB angular power spectrum against Planck 2018 data. The dashed lines represent $\LCDM$ while the dotted lines correspond to the best fit $\fTCDM$ model.  They correspond to the best-fit values of the model parameters given in Table \ref{Table:Results}.
    The $\fTCDM$ model slightly noticeably deviates from $\LCDM$
    only in the case of $D^{TT}_\ell$, and only for low multipoles}.}
    \label{Fig:CMB_angular}
\end{figure*}
\begin{figure*}
    \centering
    \includegraphics[scale=0.28]{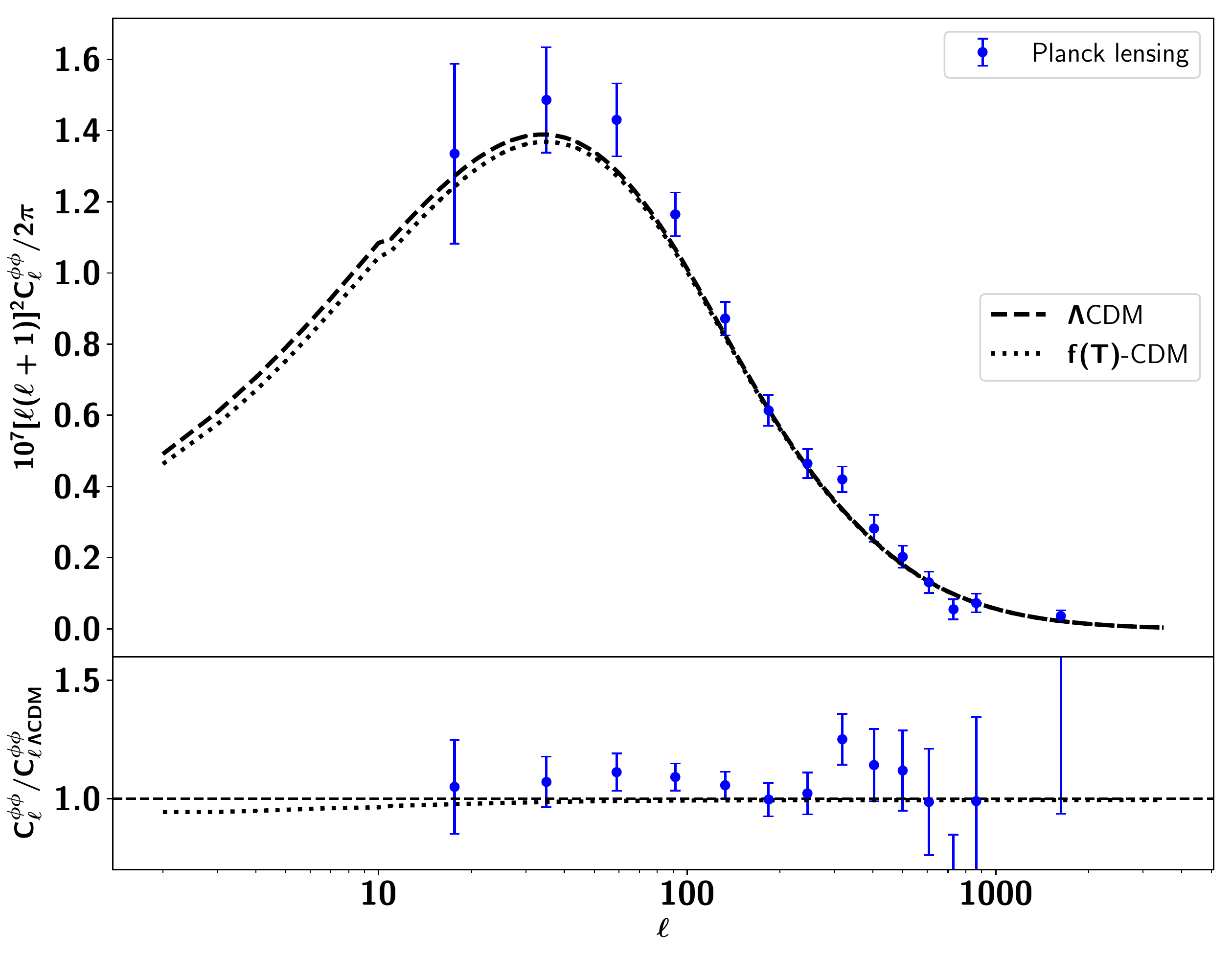}
    \includegraphics[scale=0.28]{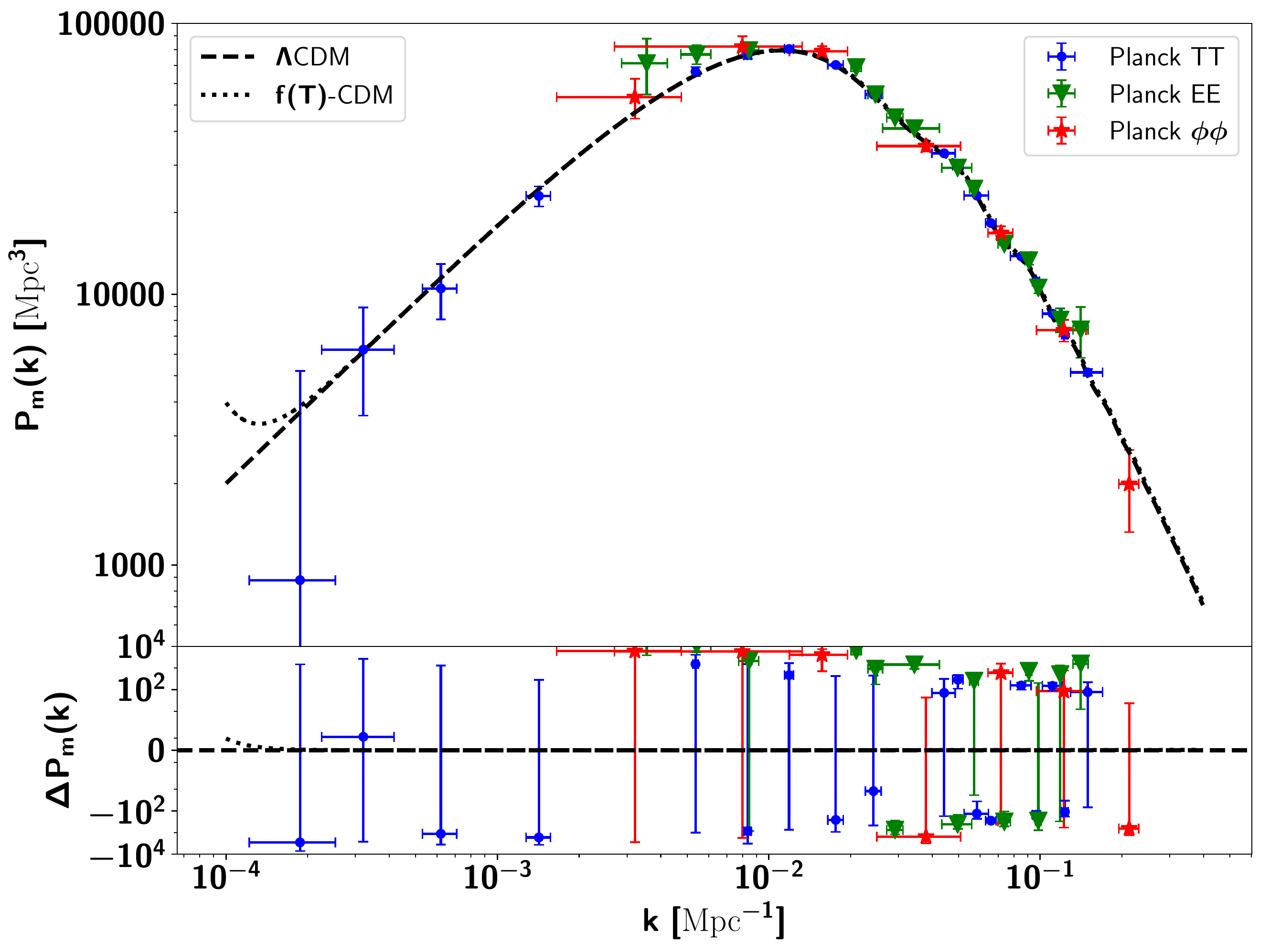}
    \caption{Same as in Fig.~\ref{Fig:CMB_angular} but for the CMB lensing-potential power spectrum (left panel), and the matter power spectrum extrapolated to $z=0$ (right panel) whereas $\Delta P_m= P_m^{\fTCDM}-P_m^{\LCDM}$.}
    \label{Fig:CMB_lensing}
\end{figure*}

The CMB angular power spectra are sensitive to the scalar fluctuations, which are determined by the evolution of the gravitational fluctuations $\phi$ and $\psi$. In the particular case of the $f(T)$ scenario these are given by Eqs. \eqref{eq:pert1} and \eqref{eq:pert2}.

In Fig. \ref{Fig:CMB_angular}, using the best-fit values of the cosmological parameters in Table \ref{Table:Results}, we plot the CMB (TT, EE and TE) power spectra for the $\LCDM$ and $\fTCDM$ together with the corresponding PL18 power spectra derived from two-point correlation functions of Planck CMB maps \citep{Aghanim:2018eyx}. As may be expected, given that our modified gravity model reduces to GR to high accuracy at
redshifts relevant to recombination era, the corresponding $f(T)$ spectra are virtually indistinguishable from the standard scenario;
notably the acoustic peaks of the CMB for both have the same amplitudes and locations.
Only the discernable difference is at low-$\ell$ multi-pole ($\ell \lesssim 30$). This is similar to those obtained from phantom dark energy, c.f. \cite{Alestas:2020mvb}. As the slight difference occurs in a regime where it is swamped by cosmic
variance, the model is practically indistinguishable from $\LCDM$ from the perspective of the CMB angular power spectrum.

\subsection{CMB lensing power spectrum}\label{Sec:CMB-lensing}
In Fig. \ref{Fig:CMB_lensing}, left panel, we plot the CMB ($\phi\phi$) power spectrum for the $\LCDM$ and $\fTCDM$ together with the corresponding PL18 power spectra derived from four-point correlation function of Planck CMB maps \citep{Aghanim:2018eyx}. Again  only slight deviations from $\LCDM$ and on scales where cosmic variance is large.
\subsection{Matter power spectrum at $z = 0$}
\label{Sec:Matter-PS}

In Fig.~\ref{Fig:CMB_lensing}, right panel, we show the linear matter power spectrum (at $z=0$) for the $\LCDM$ and $\fTCDM$, together with the corresponding PL18 (TT, EE and $\phi\phi$) power spectra \citep{Aghanim:2018eyx}. We do not include datasets from galaxy surveys, since the agreement between $f(T)$ and $\LCDM$ at most  scales are clear enough. We therefore do not
expect the $f(T)$ model to be constrained by current
large-scale structure observations.

On the other hand,  on very large scales ($k < 5\times 10^{-3}$),
significant differences are present. Notably, the $f(T)$  matter power spectrum $P_m (k)$ does not vary on large scales
as  $\sim k^4$ as in $\LCDM$. As we will see below (Section \ref{Sec:Lin-pert-evol}) these deviations are dominated by a gravitational slip term. They occur for scales quite near the horizon ($k\sim 2.4\times 10^{-4}$ Mpc$^{-1}$), and their
origin can be understood as follows.  As noted in
Section~\ref{Sec:Linear-pert}, the modification of gravity that our $f(T)$ model
entails principally affects superhorizon
scales. And since the theory is virtually indistinguishable from GR unless the dark energy
contribution is significant, only modes that enter the horizon after $z \lesssim 1$ are affected.
In the present theory the discrepancies occur at scales where cosmic variance rules out any
practical constraint on the associated cosmology. This effect may in principle nevertheless constrain
theories where the divergence from GR is present earlier (including any
early dark energy model invoking $f(T)$ gravity).

\subsection{Evolution of the matter power spectrum relative to the standard scenario}\label{Sec:Lin-pert-evol}
\begin{figure*}
    \centering
    \includegraphics[width=0.5\textwidth]{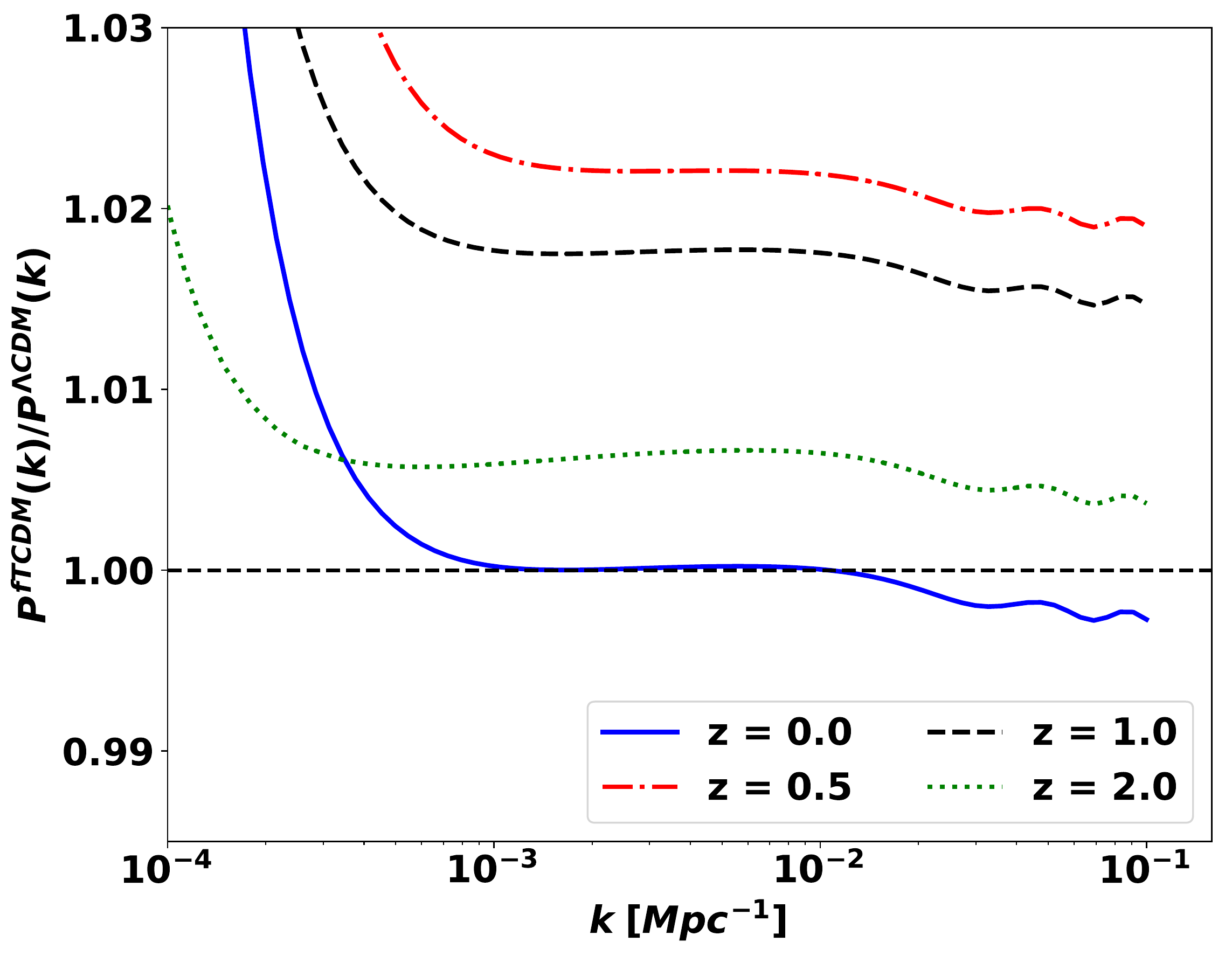}
    \caption{Ratio of matter power spectrum at different redshifts.}
    \label{Fig:MattPS_ratio}
\end{figure*}

In general, deviations in the evolution of linear perturbations in a modified gravity model
can differ from standard scenario due to three different effects:  modified background expansion;
the rescaling of the Newtonian constant at linear scales (in our case through  a scale independent $Q(a)=1/f_T$); or due to a scale and time dependent gravitational slip $\zeta(a,k)$.
In the following we wish to disentangle these effects, and delineate their consequences on the
time evolution of the power spectrum.

{In Fig. \ref{Fig:MattPS_ratio},
We plot the ratio of the matter power spectrum with respect to $\LCDM$ at various redshifts. At
scales $k \gtrsim 10^{-3}$ Mpc$^{-1}$,  relevant to BAO and
the amplitude of the growth of the structure $\sigma_8$, the ratio increases systematically
in the range $0.5 < z < 2$, before decreasing to approximately match the $\LCDM$ at $z=0$. This
effect is related to the modified background expansion.
The growth rate is enhanced at redshifts greater
than some critical value $z_c \sim 0.5$,
because the
expansion rate $H(z)$ is lower in the $f(T)$ phantom regime than in $\LCDM$
(cf.  Fig. \ref{Fig:H0-tension}, right panel; and Fig.~\ref{Fig:ext_data}, upper right panel).
The $f(T)$ Hubble rate then
crosses the corresponding  $\LCDM$ rate at $z_c$,  becoming larger for  $0 \leq z < z_c$ (including
a larger final $H_0$ value, which is necessary if the distance to the CMB is to be kept constant; see also \citep{El-Zant:2018bsc}).
During this latter epoch,  the growth rate is suppressed}.

\begin{figure*}
    \centering
    \includegraphics[width=\textwidth]{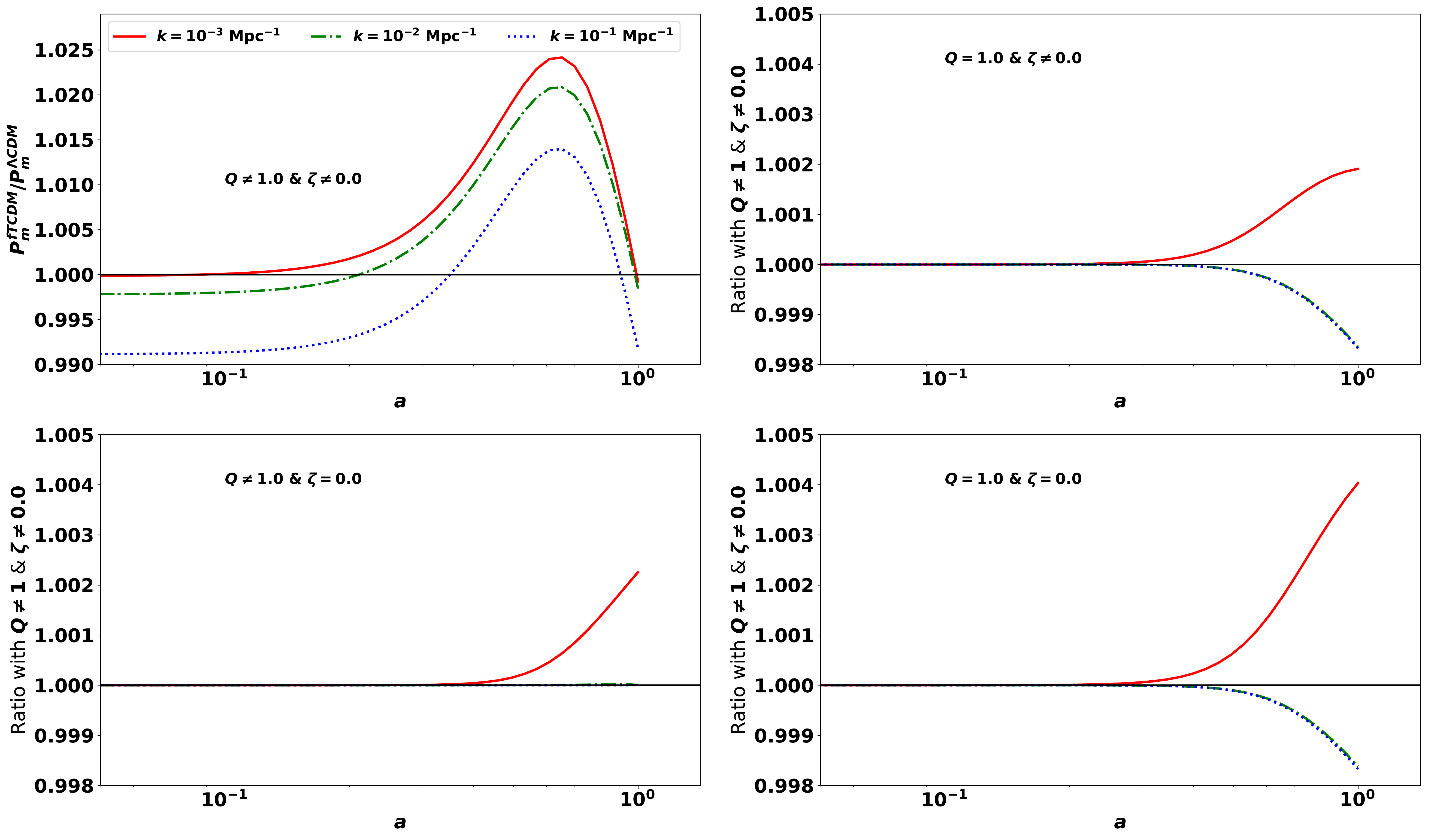}
    \caption{Top left panel: evolution, with scale factor,
    of the ratio of the matter power spectrum of our best fitting $f(T)$ model relative
    to the standard scenario. The other plots involve corresponding calculations performed
    with the same $f(T)$ model driving the background expansion but with the rescaling of Nerwtonian
    constant or the gravitational slip turned off ($Q = 1$ or $\zeta = 0$, respectively). The lines show the ratios of the results obtained thus relative to the ones shown
    in the upper left panel. The small deviations show that, at the scales considered here, the
    deviations from the standard scenario are primarily due to the different background expansion.}
    \label{Fig:Delta_sub}
\end{figure*}

\begin{figure*}
    \centering
    \includegraphics[width=0.5\textwidth]{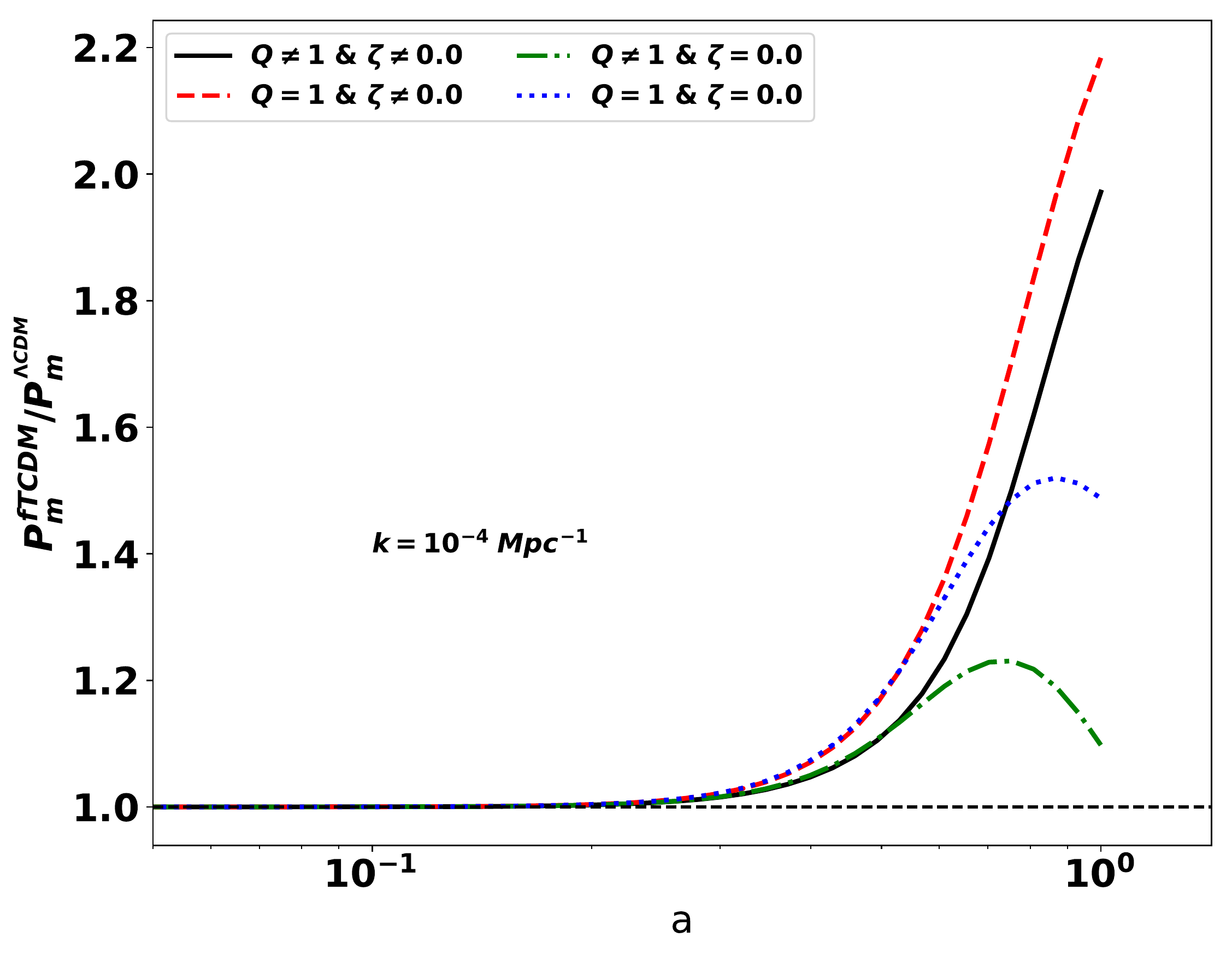}
    \caption{Ratios of the power spectra of the best fit $f(T)$ model
    to that of the standard scenario, with
    the modified gravity parameters alternatively turned on and off ($Q =1$ or $\zeta=0$),
    illustrating the importance of the
    gravitational slip term in modifying the spectrum on the largest scales.}
    \label{Fig:Delta_super}
\end{figure*}

Only on scales approaching the horizon ($k < 10^{-3}$ Mpc$^{-1}$), do discrepancies from
to the standard scenario become large, and directly driven by the
modified gravity; namely the gravitational slip.
To illustrate this,  we plot the evolution of
the ratio of the power spectrum
relative $\LCDM$ in the upper left panel
of Fig.~\ref{Fig:Delta_sub}, for values of $k$ significantly smaller
than the horizon. In order to distinguish the contribution of the rescaling of the gravitational strength $Q$, and the gravitational slip $\zeta$, we repeat the calculations in the upper left panel
with the same with either the former ($Q=1$) and/or the latter effect
turned off ($\zeta =0$).
We then divide the results thus obtained by the corresponding ones in the upper left panel
of Fig.\ref{Fig:Delta_sub}. The  results are shown in the other three panels.
As can be readily inferred, the modification are minor
(at most on the level  of a few parts in a thousand).
It is only quite close to horizon scales that major differences are present. To
illustrate the point, we plot the corresponding results for
$k = 10^{-4}$ Mpc$^{-1}$ (Fig.~\ref{Fig:Delta_super}).

\section{Conclusion}\label{Sec:Conclusion}

The cause of late time cosmic acceleration remains a major conundrum.
Besides a cosmological constant or dynamical dark energy,
it is possible to source the acceleration
through modifications of gravity at the scales
associated with dark energy.
Such modifications will also  affect
the evolution of large scale linear perturbations modes.
For these perturbations are small during the late accelerated
expansion phase, and thus in the regime where their growth can be
touched by the modified gravity. Such effects can in principle serve to constrain and
distinguish these theories from dynamical dark energy models.
In general, a theory that is empirically viable
at the level of the background
expansion dynamics, should also
not affect the growth of perturbations in ways that contradict
the large array of precise measurements
on linear scales.

We have extended to the linear perturbation regime
our study of an exponential modified
teleparallel gravity. In its context, late time acceleration as torsion scalar $T$ tends to a value $T_0 \sim H_0^2$ on cosmological (background) scales at late times. The model
tends to the teleparallel equivalent to GR (where $T$ plays the role of the Ricci scalar)
at earlier times. Before reaching the ``cosmological constant'' stage (at a future fixed point $T \rightarrow T_f=-12\beta H_0^2$ as $z\to -1$, see Sec. 3.2.3 in paper I \citep{Hashim:2020sez}),
the background dynamics displays phantom-like behaviour, while circumventing problems related
to the null energy condition that must be overcome when such a regime
is invoked via dynamical dark energy.

We derive the full CMB and matter power spectra, and show that the deviations
from their counterparts arising in the standard $\LCDM$ scenario
are generally quite minor --- at the percent level or below ---
and primarily associated with the different background expansion.
Only at late times, and close to horizon scales, are the modifications
to the gravitational field important enough to produce significant
changes to the power spectrum of matter perturbations. These are swamped
by cosmic variance in the current theory, but may in principle be used
to constraint other models where the modifications come earlier or have
different form, particularly any early dark energy models.

Shortly after the discovery of cosmic acceleration, the
concordance model, assimilating early and late universe measurements,
now more commonly known as $\LCDM$, ascended.
However, increasingly precise data revealed apparent tensions
between late and early (CMB-inferred) universe parameter estimates, notably of the current
Hubble expansion rate. We confirm here that, seen solely as a tension between the CMB and direct measurements, this tension may be successfully alleviated by late time phantom-like dark energy,
such that exhibited by our modified gravity model.

Indeed, using the full CMB spectrum, we find that the  Planck (TT+TE+EE+lensing) base-$\fTCDM$ alone predicts Hubble constant $H_0=72.24\pm 0.64$ km/s/Mpc which  alleviates the $H_0$ tension with R20+H0LiCOW to be at 0.8$\sigma$ level, while in $\LCDM$ it reaches 4.8$\sigma$.
Furthermore, the number of free parameters in both models are the same. Including BAO distance data however
decreases the expected value of the Hubble constant in the context the best fit exponential $f(T)$ model to
$H_0=71.49\pm 0.47$ km/s/Mpc (Table \ref{Table:Results}). The effect leading to this decreased
value moreover reflects systematic deviations at the redshifts probed by
BAO measurements. In fact, although the current theory is viable --- in the sense that it fits the data with  $\chi^2$ values comparable to $\LCDM$ --- it shows systematic deviations with not only the BAO but also the SNIa distances and the growth of structure ($f\sigma_8$), where it is in more severe tension than the standard model
(Fig. \ref{Fig:ext_data}).
Higher statistical weight to BAO data and more accurate measurements of SNIa distances could make such discrepancies even more severe.
As, unlike other viable $f(T)$ theories, the present theory does not reduce to $\LCDM$ for some choices of extra free parameters,  the  aforementioned trends are generic, and should apply to any modified gravity or dynamical dark energy model with phantom behaviour.

We conclude that modified gravity theories
involving extensions of the telparallel formalism can support viable cosmologies,
both in terms of the background dynamics as well as at the linear perturbations level,
without introducing extra parameters. Furthermore,
the phantom-like behaviour that the current theory
entails may entirely alleviate the apparent $H_0$ tension, if this is seen solely as a tension between
the CMB inferred values and direct local measurements. But,  as expected in any scenario
invoking such behaviour, other tensions arise  when additional constraints on the way to the CMB,
such as BAO distance measurements, are encountered.

\subsection*{Acknowledgements}

WEH would like to thank Eleonora Di Valentino for useful discussions about the ultra-conservative estimate of $H_0$ in comparison to the value obtained in this work, and also partial discussion about the KV-450 analysis. The likelihood analysis presented in this work were done on the Sciama High Performance Compute (HPC) cluster which is supported by the ICG, SEPNet and the University of Portsmouth. We acknowledge Vivien Bonvin and Martin Millon for providing the publicly available codes that are used to make Fig.~\ref{Fig:H0-tension} (left panel)
\href{https://github.com/vbonvin/H0_tension}{\faGithub} \url{https://github.com/vbonvin/H0_tension}.
This project was supported financially by the Science and Technology Development Fund (STDF), Egypt. Grant No. 25859.

\appendix
\section{Distance measures used}
\label{App:Definitions}

We provide here definitions for the supernovae and BAO distance measures used
in Section \ref{Sec:Data-results}.

The supernovae distance modulus is
\begin{equation}\label{dist_mod}
\nonumber \mu(z)=m(z)-M =25+5\log _{10}{D_{L}(z)},
\end{equation}
where $M$ denotes the absolute magnitude, and the luminosity distance $D_L$ measured in Mpc
\begin{equation}\label{luminosity_distance}
    D_L(z)=\frac{(1+z)}{H_0}\int_0^z \frac{dz'}{E(z')}, \quad E(z)=H(z)/H_0.
\end{equation}

The angular distance
\begin{equation}\label{ang_dist}
D_{A}(z)=\frac{1}{(1+z)H_0}\int_0^z \frac{dz'}{E(z')}.
\end{equation}
The volume averaged comoving angular diameter distance
\begin{equation}\label{DV}
    D_V(z)=\left[D_M^2\frac{cz}{H(z)}\right]^{1/3},
\end{equation}
where $D_M$ is the comoving angular distance,
\begin{equation}\label{comoving_angular_distance}
    D_M(z)=(1+z)D_A(z).
\end{equation}
The sound horizon at drag epoch $z_d$
\begin{equation}\label{sound_horizon}
    r_d=\int_{z_d}^\infty \frac{c_s(z)}{H(z)}dz,
\end{equation}
where $c_s(z)$ is the baryons (with density $\rho_b$) and photons (with density $\rho_\gamma$) speed of sound
\begin{equation}\label{sound_speed}
    c_s(z)=\frac{1}{\sqrt{3}}\left[1+\frac{3\rho_b}{4\rho_\gamma}\right]^{-1/2}.
\end{equation}

\section{$S_8$ tension}\label{App:S8-tension}
We compare Planck constraints on $S_8 = \sigma_8 \sqrt{\Omega_m/0.3}$ in context of $\LCDM$ with those within the  $\fTCDM$ model. We use the cosmic shear measurements from KiDs+VIKING-450 (KV-450), and reproduce the fiducial likelihood run, as in \citep{Hildebrandt:2018yau} for the exponential IR $f(T)$ theory.

\begin{figure}
    \centering
    \includegraphics[width=0.49\textwidth]{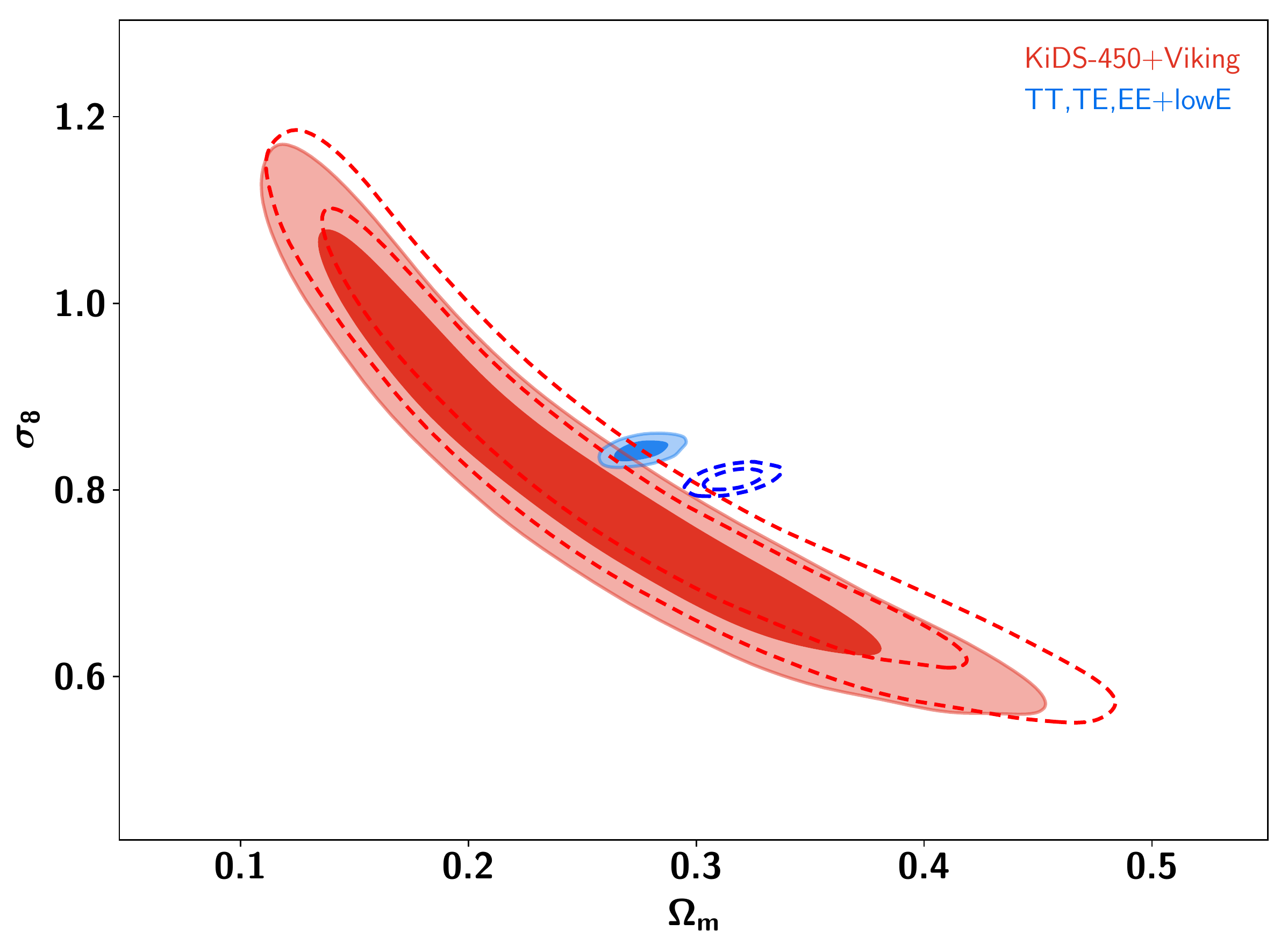}
    \includegraphics[width=0.49\textwidth]{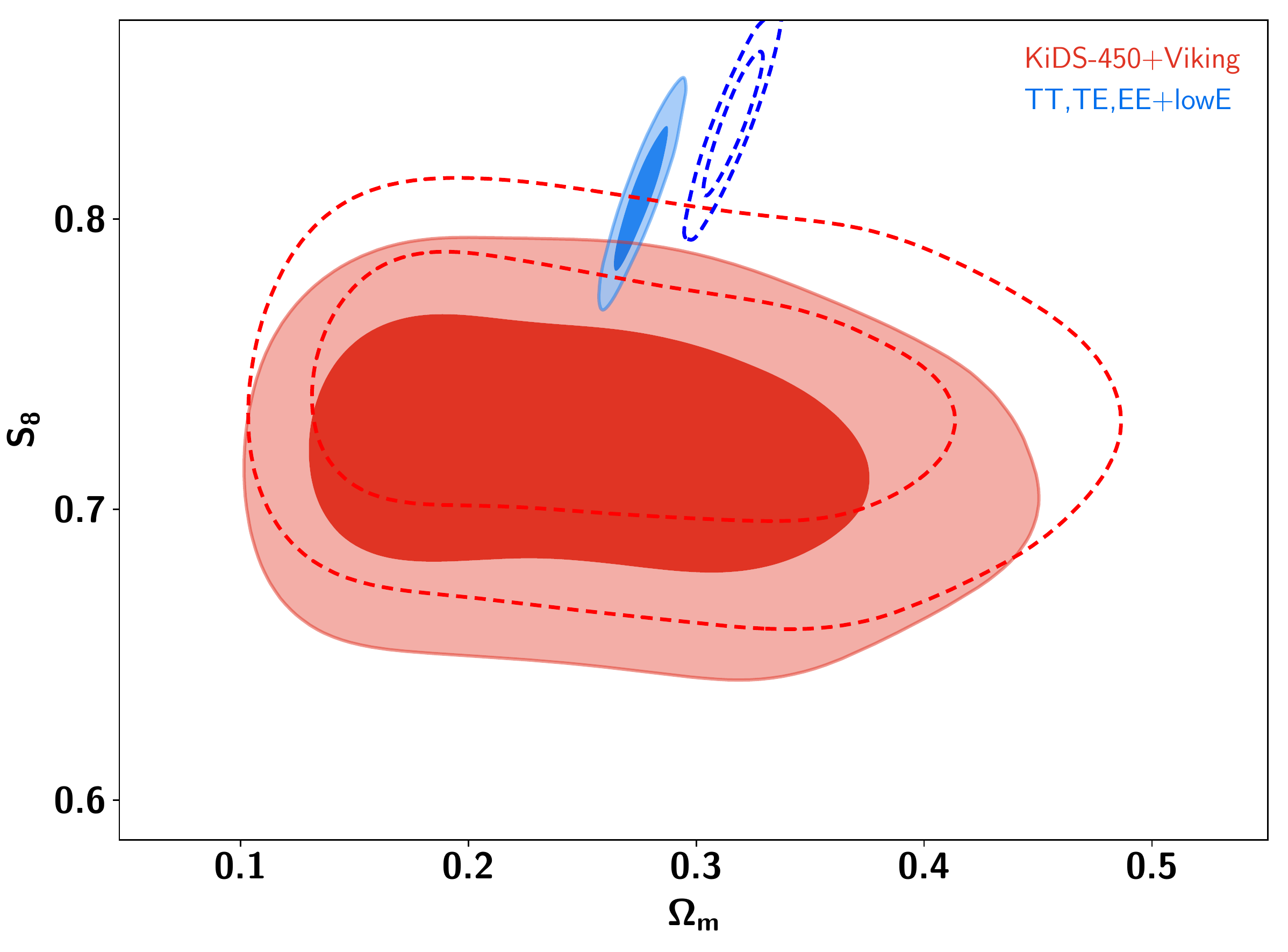}
    \caption{{68\% and 95\% CL in the $\sigma_8-\Omega_m$ plane (upper panel) and $S_8-\Omega_m$ plane (lower panel). Dotted contours are for $\LCDM$ and solid contours are for $\fTCDM$, while red color is for KV-450 and blue color is for Planck (TT,TE,EE+lowE)}.}
    \label{Fig:S8-tension}
\end{figure}

In the context of the current $\fTCDM$ theory, modifications to the linear power spectrum
are significant at late time only near horizon scales ($k < 10^{-3}$ Mpc$^{-1}$). Otherwise
the modifications are small ( $\lesssim 1.5$\%) and nearly scale free, as they almost entirely
arise from background evolution change (see Sec. \ref{Sec:Lin-pert-evol}). In this context, the VK-450 fiducial likelihood can be used to analyze the exponential IR $f(T)$ theory assuming no change
in the power spectrum. We also assume that the results are unaffected by evolutiom at the nonlinear level. This last assumption is
justified  both by the expectation that the IR $f(T)$ theory is expected to be even closer
to GR at smaller scales, and
by the results of KiDS-450 where the removal of the angular scales sensitive to nonlinear physics does not alter the main result \citep{Joudaki:2016kym}. We visualize the results in Fig. \ref{Fig:S8-tension} as seen in the ($\sigma_8$, $\Omega_m$) and ($S_8$, $\Omega_m$) planes.

In Table \ref{Table:Results}, the Planck base-$\LCDM$ measures $\sigma_8=0.8117\pm 0.007$, $S_8=0.833\pm 0.016$ and $\Omega_m=0.3162\pm 0.008$ (TT,TE,EE+lowE). We obtain VK-450 results of the same set of parameters as \begin{equation*}
\left.
\begin{array}{ll}
   \sigma_8 = & 0.82^{+0.11}_{-0.20},\\[5pt]
   \Omega_{m} = & 0.268^{+0.067}_{-0.12},\\[5pt]
   S_8 = & 0.737^{+0.033}_{-0.030}.
  \end{array}
\right\}\hbox{68\% KV-450 ($\LCDM$).}
\end{equation*}
We note that the cosmic shear measurements of $S_8$ is at $\sim 2.6\sigma$ tension with our obtained CMB base-$\LCDM$ value.

We next repeat the same analysis for the exponential IR $f(T)$ theory. In Table \ref{Table:Results}, the Planck base-$\fTCDM$ measures $\sigma_8=0.8425\pm 0.007$ which is higher than the $\LCDM$ value, $S_8=0.808\pm 0.016$ and $\Omega_m=0.2758\pm 0.007$ (TT,TE,EE+lowE). Also, We obtain VK-450 results of the same set of parameters as
\begin{equation*}
\left.
\begin{array}{ll}
   \sigma_8 = & 0.82^{+0.11}_{-0.18},\\[5pt]
   \Omega_{m} = & 0.252^{+0.062}_{-0.100},\\[5pt]
   S_8 = & 0.719^{+0.033}_{-0.030}.
  \end{array}
\right\}\hbox{68\% KV-450 ($\fTCDM$).}
\end{equation*}
We find that the cosmic shear measurements of $S_8$ using VK-450 dataset is at $\sim 2.4\sigma$ tension with our obtained CMB base-$\fTCDM$ value, which reduces the $S_8$ tension by $0.2 \sigma$ due to lower $\Omega_{m}$ value in comparison with $\LCDM$ although the tension in $\sigma_8$ has been increased.

\bibliographystyle{JHEP}

\providecommand{\href}[2]{#2}\begingroup\raggedright\endgroup

\end{document}